\documentclass[10pt,conference,compsoc,a4paper,twocolumn]{IEEEtran}\usepackage[]{graphicx}\usepackage[]{color}
\makeatletter
\def\maxwidth{ %
  \ifdim\Gin@nat@width>\linewidth
    \linewidth
  \else
    \Gin@nat@width
  \fi
}
\makeatother

\definecolor{fgcolor}{rgb}{0.345, 0.345, 0.345}

\usepackage{framed}
\makeatletter
 {\par\unskip\endMakeFramed%
 \at@end@of@kframe}
\makeatother

\definecolor{shadecolor}{rgb}{.97, .97, .97}
\definecolor{messagecolor}{rgb}{0, 0, 0}
\definecolor{warningcolor}{rgb}{1, 0, 1}
\definecolor{errorcolor}{rgb}{1, 0, 0}
\newenvironment{knitrout}{}{} %

\usepackage{alltt}

\usepackage[utf8]{inputenc}
\usepackage[english]{babel}

\usepackage[font=small]{caption}
\usepackage{subcaption}
\usepackage{multirow}

\usepackage{graphicx}

\usepackage{tikz}
\usepackage{tikzscale}
\usetikzlibrary{positioning,shapes,fit,calc,arrows,arrows.meta}

\usepackage{MnSymbol}

\usepackage{pifont}
\newcommand{\yes}{\ding{51}}

\newcommand{\yesbut}{(\yes)}

\usepackage{csquotes}

\usepackage{tabularx}
\usepackage{ragged2e}
\newcolumntype{P}[1]{>{\RaggedRight\arraybackslash}p{#1}}
\newcolumntype{C}[1]{>{\centering\arraybackslash}p{#1}}

\usepackage[style=ieee,sorting=nyt]{biblatex}
\DeclareSourcemap{\maps[datatype=bibtex]{\map{
    \step[fieldsource=booktitle, match={Conference}, replace={Conf.}]
    \step[fieldsource=booktitle, match={Symposium}, replace={Symp.}]
    \step[fieldsource=booktitle, match={Proceedings of the}, replace={Proc.}]
    \step[fieldsource=booktitle, match={Proceedings}, replace={Proc.}]
    \step[fieldsource=booktitle, match={International}, replace={Intl.}]
    \step[fieldsource=journal, match={International Journal}, replace={Intl. J.}]
}}}
\DeclareFieldFormat[book,inbook,incollection,inproceedings]{series}{#1}

\usepackage{hyperref}
\usepackage{xcolor}
\definecolor{shadecolor}{HTML}{990000}

\definecolor{TFFrameColor}{HTML}{D6D6D6}
\definecolor{TFTitleColor}{HTML}{000000}
\hypersetup{
    colorlinks,
    linkcolor={red!60!black},
    citecolor={green!50!black},
    urlcolor={blue!95!black}
}

\usepackage[capitalize]{cleveref}

\newcommand{\subparagraph}{}
\usepackage{titlesec}
\titlespacing*{\section}{0pt}{3.5ex plus 1ex minus .2ex}{2.3ex plus .2ex}
\titlespacing*{\subsection}{0pt}{3.25ex plus 1ex minus .2ex}{1.5ex plus .2ex}
\titleformat{\subsubsection}[runin]{\bfseries}{}{0em}{}[\mbox{\hspace{1em}}]

\usepackage{titletoc}
\titlecontents{section} [7mm]%
    {\addvspace{0mm}}%
    {\bf\contentslabel{6mm}}%
    {\hspace{-7mm}\bf}%
    {\bf\titlerule*[2.7mm]{.}\contentspage}%
    [\addvspace{0mm}]%
\titlecontents{subsection} [15mm]%
    {\addvspace{0mm}}%
    {\contentslabel{8mm}}%
    {}%
    {\titlerule*[2.7mm]{.}\contentspage}%
    [\addvspace{0mm}]%

\usepackage{enumitem}
\setlist{%
    leftmargin=1.5em%
}

\usepackage{framed}

\newcommand{\sRef}[1]{{\bfseries\hyperref[session:#1]{#1}}}

\newcommand{\cRef}[1]{{\bfseries\sffamily\hyperref[company:#1]{#1}}}

\newcommand{\NA}{{\sffamily\itshape n.a.}}
\IfFileExists{upquote.sty}{\usepackage{upquote}}{}
\begin{document}

\title{PP-ind: Description of a Repository\\ of Industrial Pair Programming Research Data}
\author{%
    \IEEEauthorblockN{Franz Zieris}
    \IEEEauthorblockA{zieris@inf.fu-berlin.de\\Freie Universtität Berlin\\Berlin, Germany}
    \and
    \IEEEauthorblockN{Lutz Prechelt}
    \IEEEauthorblockA{prechelt@inf.fu-berlin.de\\Freie Universtität Berlin\\Berlin, Germany}
}

\maketitle
\thispagestyle{plain}
\pagestyle{plain}

\begin{abstract}
    PP-ind is a repository of audio-video-recordings of
    industrial pair programming sessions.
    Since 2007, our research group has collected data in
    13 companies.
    A total of 57 developers worked together
    (mostly in groups of two, but also three or four)
    in 67 sessions
    with a mean length of 1:35 hours.
    In this report, we describe how we collected the data and provide summaries
    and characterizations of the sessions.
\end{abstract}

\tableofcontents

\section{Introduction}

Pair programming (PP) is a software development practice in which two developers
work closely together on a technical task on the same computer.
It was popularized by Kent Beck who sees it as the central practice of
eXtreme Programming and describes it as \emph{``a dialog between
to people trying to simultaneously program (and analyze and design
and test) and understand together how to program better''} \cite[p.~100]{Beck99}.

Controlled experiments on pair programming have shown mere tendencies in
terms of effects on quality and effort with much variation left to be explained
\cite{HanDybAri09}.
In the words of the authors of a large experiment with almost 300 hired
consultants:
\emph{``we are still far from being able to explain why we observe the given
effects''} \cite{AriGalDyb07}.

Our research group has been collecting industrial pair programming sessions
since 2007.
We record pair programming as it happens ``in the wild'' in order to
understand how it actually works and what really matters in everyday practice.
In particular, we record the pairs' converstation, their screen content,
and a webcam video showing their gestures and posture.

This kind of data data is amendable to different types of analyses.
We describe our qualitative approach in \cite{SalPloPre08,SalPre13-baseconbook}.
In this report, we describe the technicalities of how we collected the data
and provide some metadata for each session.
Several researchers have contributed a lot of time to collecting and
processing that data, and we want to give credit.
The raw data itself cannot be released to the public because of non-disclosure
agreements with the respective companies.
As a proxy, we characterize the companies, the developers, and their
PP sessions.

This report is structured as follows:
We discuss our fundamental approach to collecting empirical data on pair
programming (Section~\ref{sec:fundamental}) and describe our generic
data collection protocol (Section~\ref{sec:protocol}).
We introduce some terminology and describe the structure of our data
(Section~\ref{sec:terminology-structure}).
We give an overview of our repository (Section~\ref{sec:overview}) and
then discuss the individual contexts and cases (Section~\ref{sec:repo}).
We close with a discussion of the properties and limitations of our
data collection (Section~\ref{sec:discussion}) and an overview
of which data has been used in which publications so far
(Section~\ref{sec:publications}).
In Appendix~\ref{sec:technicalities}, we explain the technical details
of how we record and process PP sessions.

We provide repository meta-data, partial transcripts, questionnaires,
and additional material as a public data set \cite{PPind-meta}.

\section{Fundamental Considerations}
\label{sec:fundamental}

There are two fundamental considerations to our data collection:
First, we record pair programming as it happens in industry.
Second, we consider the pair programming session as the basic unit.
Both considerations were driven by our research interests.

\subsection{Naturalistic Industrial Setting}

Our research wants to achieve practical relevance.
Therefore, we study industrial settings with
\emph{professional software developers}
working on their \emph{everyday tasks}.
This also entails that the developers work in their normal development
environment, with partners they chose to work with,
at times and to an extent they decide themselves.

We primarily rely on \emph{observation} of developers working in pairs,
as opposed to interviews.
To enable a thorough analysis, we \emph{record} the pair programmers.
In particular, the pair members' interactions with one another and their
computer(s) as well as the contents of their screen(s) need to be captured
in audio and video.
The necessary recording infrastructure somewhat reduces the naturalism of the
observed session; we discuss the effects in
\cref{sec:recording-effects}.

\subsection{The Pair Programming Session as a Unit}

Our data collection starts when the developers have already made
the decision to work as a pair.
Their decision, just as the project they work in, the task(s) they work on,
their software system, and their team structure all may ``echo'' in their
session and so knowing these things can be helpful for understanding
their activities---but it is not an important goal of our data collection.

\section{Data Collection Protocol}
\label{sec:protocol}

In our research group, Stephan Salinger and Laura Plonka initiated our
industrial data collection efforts and they devised a protocol that
served as the basis for collecting data in all companies.

The data collection protocol is \emph{generic} in two ways.
First, it is adapted in each particular installment at a company on-site to deal
with constraints, to seize opportunities, and to fit the particular research
focus of the researcher (see Section~\ref{sec:contexts}).
Second, the protocol is still more or less independent from any particular
research question regarding pair programming, as the resulting data can be
reused for different purposes (some conditions apply, which we discuss in
\cref{sec:discussion}).

\subsection{Protocol Overview}

After a company has been approached and probed whether the company would be
open to have some of their programming sessions recorded,
the overall research goal, the procedure, extent, and purpose of the main data
collection are explained in a presentation for the development teams.
We explain that all participation is voluntary and that their
individual agreement to be recorded can be revoked at any point
during a session.
These are the steps for each session recording:
\begin{itemize}
    \item
    After a pair announces that it is willing to have their next pair session
    recorded,
    the recording infrastructure is set up. The \textbf{session recording} is
    started once the developers are ready
    (see \cref{sec:method-session-recording} for details).

    \item
    Optionally, both developers fill out \textbf{questionnaires} before and/or
    after their session, in which the developers state their names, development and
    pair programming experience, characterize the nature of their task, and
    whether it went as they intended
    (see \cref{fig:questionnaires}).

    \item
    Afterwards, the researcher does a \textbf{quick analysis} of the material
    during which she looks for peculiarities that catch her attention.
    The main purpose of this step is to inform the next activity.

    \item
    The researcher then conducts a \textbf{reflective interview} with the
    developers on the day after the recording.
    This activity serves to collect background
    information and providing developers with feedback in return for their
    agreement to have their PP session recorded and scrutinized.
    These interviews are audio-recorded.
\end{itemize}

\subsection{Recording Sessions}
\label{sec:method-session-recording}

The software developers themselves decide when and for how long they want to
have their work recorded.
They work on their own machines, in their normal environment, on their everyday
tasks, and with the partner they chose.

\begin{figure*}
    \includegraphics[width=\linewidth]{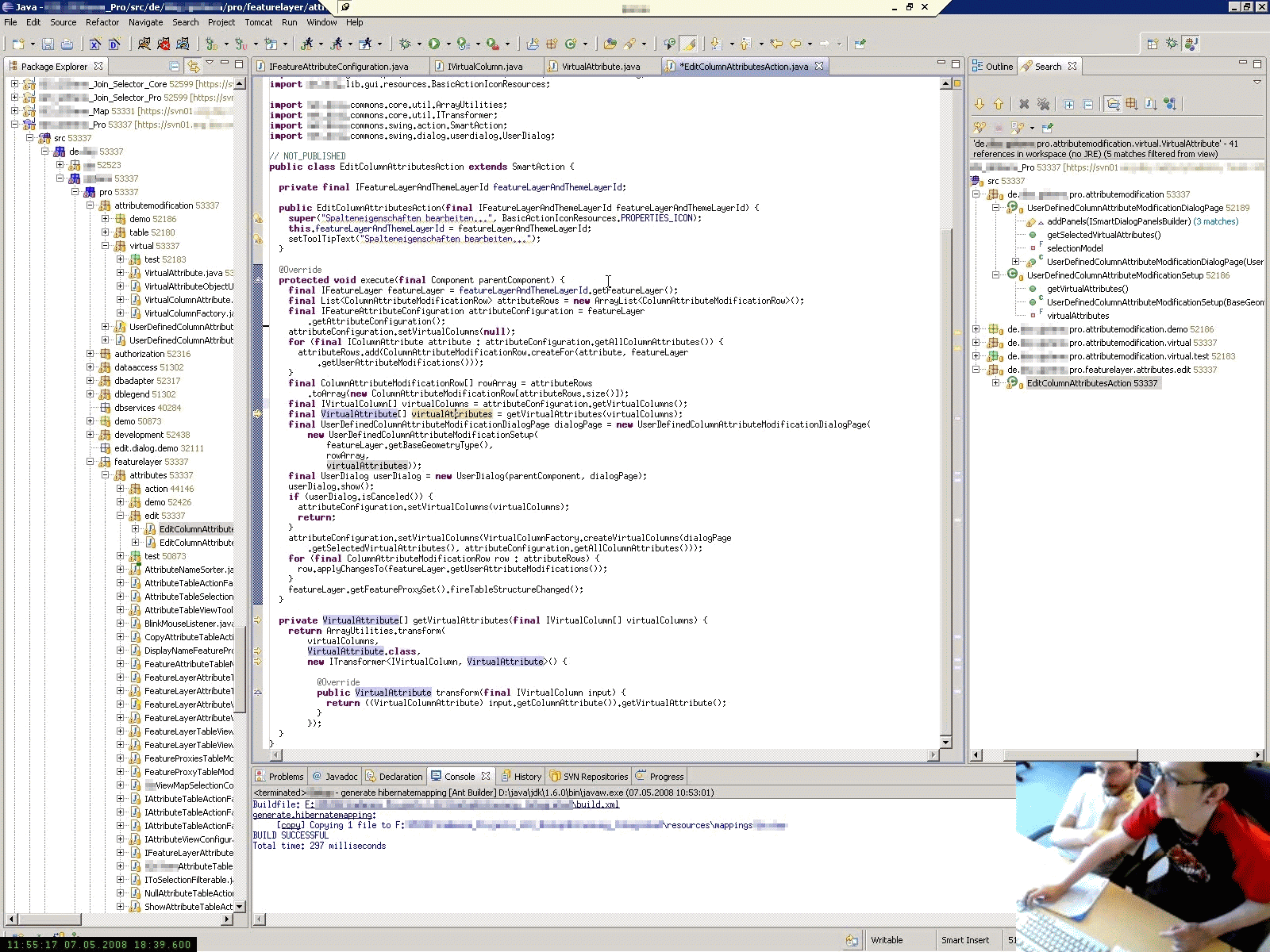}
    \caption[Still frame of a session recording]{%
        Still frame from the \sRef{CA2} session recording
        with screencast in the background and the webcam video layered on top.}
    \label{fig:video-still-frame}
\end{figure*}

The session recordings as technical artifacts consist of
a screencast of the pair's monitor(s),
the pair's conversation as audio,
and a webcam video showing the two pair members' interaction.
These three sources are combined to a self-contained video file as
illustrated in \cref{fig:video-still-frame}.
Both webcam feed and screencast are captured at 5 to 15 frames per
second (depending on hardware capabilities),
which is enough to distinguish
individual keystrokes, to follow mouse movements, and the see the developers'
gestures.
The final video resolution depends on the developers' display(s) and
recording setup and ranges from 1024$\times$768 to 2560$\times$1440 pixels.

The recording process relies on one of three generations of hardware and
software components.
General setup:
The developers work on one machine, and screencast and webcam feed are
transmitted to another machine where they are recorded;
we explain the details in Appendix~\ref{sec:technicalities}.
The most relevant difference is that generation~1 is an unattended
recording which the researcher only gets to see once the pair is done,
while generations~2 and 3 are an online recording
which allows the researcher to also watch the session live and start her quick
analysis).

\subsection{Per-Company Differences}
\label{sec:contexts}

With each installment of the data collection protocol at a new company,
there were slightly different sets of mutual expectations which resulted
from prior discussions with the partners and from evolved research interests
on our side.
We discriminate three groups here, because they shaped our
behavior and likely our subjects' behavior a bit differently.
Table~\ref{tab:context} then gives an overview of the \emph{individual}
contexts (and involved researchers) for each such research ``headline''.

\begin{table*}
    \centering
    \begin{tabular}{cll|llll}
        \hline
        \multicolumn{3}{c|}{\textbf{Company}} &
        \multicolumn{4}{c}{\textbf{Data Collection}} \\
        \textbf{ID} & \textbf{Application Domain} & \textbf{Country} & \textbf{Year} & \textbf{Researchers} &
        \textbf{Headline} & \textbf{Focus} \\
        \hline
        \textsf{A} & Content Management System & Germany & 2007 & Salinger & PP & -- \\
        \textsf{B} & Social Media & Germany & 2007 & Salinger \& Plonka & PP & -- \\
        \textsf{C} & Geo-Information System & Germany & 2008 & Plonka & PP & workshop, PP roles \\
        \textsf{D} & Customer Relationship Management & Germany & 2008 & Plonka & PP & workshop, PP roles \\
        \textsf{E} & Logistics and Routing & Germany & 2008 & Plonka & PP & workshop, PP roles \\
        \textsf{F} & Email Marketing & Germany & 2008 & Plonka & PP & workshop, PP roles \\
        \textsf{J} & Data Management for Public Radio Broadcast & Germany & 2013 & Schenk & PP & distributed PP \\
        \textsf{K} & Real Estate Online Platform & Germany & 2013 & Salinger, Zieris & Agile \& PP & knowledge transfer
        \\
        \textsf{L} & Freelance Training Consultant & USA & 2014 & Schenk & PP & distributed PP \\
        \textsf{M} & Data Analysis in Energy and Transportation & Norway & 2014 & Zieris & PP & knowledge transfer \\
        \textsf{N} & Online Fashion Retailer & Germany & 2016 & Salinger, Schmeisky, Zieris & Onboarding & -- \\
        \textsf{O} & Online Project Planning & Germany & 2016 & Zieris & Agile \& PP & knowledge transfer \\
        \textsf{P} & Online Car Part Resale & Germany & 2018 & Zieris & PP & knowledge transfer \\
        \hline
    \end{tabular}
    \caption{%
        Contexts for sessions recorded by member of our research group.
        The research direction (``headline'' and ``focus'') was set by the
        named researchers.
        Sessions in three additional industrial contexts (\textsf{G},
        \textsf{H}, and \textsf{I}) were recorded with low technical
        quality and by students with little oversight, so important context
        information is missing.
        We exclude these low-quality sessions from our repository.}
    \label{tab:context}
\end{table*}

\begin{itemize}
    \item \textbf{PP}:
    Most companies were specifically approached with the intention to
    understand pair programming.

    While for the first contacts there was no particular focus yet,
    Plonka's \cite{Plonka12} contacts to companies \textsf{C} to \textsf{F} were
    influenced by a particular interest in the \emph{driver and navigator} roles,
    which is why the webcams in these sessions are recorded in an angle that
    shows the developers hands but occassionally cuts off their heads.

    Additionally, the data collection period was branded to these companies as a
    ``workshop'' to help developers reflect on their PP process.
    Here, one work station was set up for all pairs to use
    and the developers would put their names on a list to choose either a morning
    or an afternoon slot to be recorded for a planned maximum duration of 1.5 to 2
    hours.
    Nevertheless, once a recording was started, the developers (just as in
    all other companies) were left to work on tasks of their choosing for as long
    they wanted.

    Schenk \cite{Schenk18} was particularly interested in distributed pair
    programming and chose her contacts \textsf{J} and \textsf{L} accordingly;
    Zieris \cite{ZiePre14-ppknowtrans,ZiePre16-ppknowtrans2} is
    interested in knowledge transfer and all participants from
    companies \textsf{K}, \textsf{M}, \textsf{O}, and \textsf{P} knew that.

    \item \textbf{Agile}:
    Salinger and Zieris approached companies \textsf{K} and \textsf{O}
    in an effort to understand agile software development in general.
    According data from these contexts was collected and analyzed
    (see \cite{ZieSal13-scrumnotagile}, which is about company \textsf{K}).
    All particular PP sessions, however, were recorded exclusively for the
    purpose of understanding knowledge transfer in pair programming.

    \item
    \textbf{Onboarding}:
    Schmeisky, Salinger, and Zieris approached company \textsf{N} for
    understanding their onboarding process, i.e., how new-hires are integrated
    into the company.
    Pair programming was not a designated part of that process, but a number of
    developers agreed to be recorded while working in pairs.
    The recordings were therefore a window into the actual onboarding process,
    and were not made to understand pair programming.
\end{itemize}
Another effect of these different headlines and focuses can be seen in the
different versions of the pre- and post-session questionnaires, see
Figure~\ref{fig:questionnaires}.

\begin{figure*}
    \small
    \sffamily
    \fbox{
    \begin{minipage}[t]{0.48\textwidth}
        \textbf{Task \& Pair Items (``Pre-Session'')}
        \begin{enumerate}[leftmargin=1.3em]
            \item Task classification
            (new functionality, extend functionality, test cases, debugging,
            refactoring, or other)
            \item Short description of the task
            \item Characterization of (expected) difficulties
            \item Estimated time to completion\\
            {\rmfamily\textins{added in version 3}}
            \item Why task is worth to be worked on by a pair
            \item Professional software development experience
            \item Pair programming experience\\
            {\rmfamily\textins{added in version 2}}
            \item How well attuned to their respective partner?
            \item Expectations towards the reflective interview later on
            {\rmfamily\textins{added in version 2}}
        \end{enumerate}
    \end{minipage}
    }\hfill
    \fbox{
    \begin{minipage}[t]{0.44\textwidth}
        \textbf{Process Items (``Post-Session'')}
        \begin{enumerate}[leftmargin=1.3em]
            \item School grade for recent session
            \item Compare progress with expectations
            \item Divide session into phases
            \item Name most important phases
            \item Assess session-specific importance of each: knowledge
            transfer, developing a strategy, bug fixing, developing a design/an
            architecture, developing an algorithm, knowing an API, having the
            right idea
            {\rmfamily\textins{removed in version 2}}
            \item Points where pair constellation should have been given up
            \item Points where pair constellation was especially
            beneficial
            {\rmfamily\textins{added in version 2}}
        \end{enumerate}
    \end{minipage}
    }
    \caption{Shortened items from the pre-session and the post-session
    questionnaire which the developers filled out individually
    (except for the \textsf{D}-pairs who all handed in just one collective
    pre-questionnaire).
    There were three different versions: Version 1 was used for
    companies \textsf{A} and \textsf{B}, version 2 for 
    \sRef{CA1}, \sRef{CB1}, \sRef{CA4}, and \sRef{CA5}), and
    version 3 for \sRef{CA2} and \sRef{CA3}
    as well as for companies \textsf{D}, \textsf{E}, and \textsf{F}
    (all three questionnaire versions can be found in our data set \cite{PPind-meta}).
    Developers from companies \textsf{J} and \textsf{K} received and answered the
    \emph{task \& pair items} (except 4, 6, and 9) via e-mail \emph{after}
    their respective sessions.
    In the other installments (\textsf{M}, \textsf{N}, \textsf{O}, \textsf{P}),
    no questionnaires were used.
    Instead, the researchers asked the developers directly in the reflective
    interviews.}
    \label{fig:questionnaires}
\end{figure*}

\section{Terminology and Structure}
\label{sec:terminology-structure}

We use the following nomenclature:
\begin{itemize}
    \item
    Companies are represented by single letters: \textsf{A}, \textsf{B},
    \textsf{C}, and so on.
    \item
    Sessions are grouped by their technical context (``project'', each with
    a different set of requirements and/or different technology stack) and then
    counted up with Arabic numbers.
    \textbf{CB1} is the first recording in the second context in the
    third company.
    \item
    Developers are identified through their company and Arabic numbers,
    such as \textbf{C3}.
\end{itemize}

\subsection{Pair Programming Modes}
\label{sec:modes}

Although our primary research focus is on pair programming, we did not restrict
our recordings to a fixed setting of two developers working on the same
machine.
We distinguish three dimensions along which the collaboration modes in our
repository differ.
\begin{itemize}
    \item \textbf{Number of Developers}:
    Just as one developer may ask a colleague for help, a pair may do the same.
    Some teams even form groups of three or more people intentionally.
    So far, we recorded sessions with groups of two to four developers.

    \item \textbf{Spatial Distribution}:
    The developers can all be co-located, or all be in separate locations.
    (There could also be separate co-located subgroups,
    but we did not record such settings yet.)

    \item \textbf{Number of Active Computers}:
    The developers can share access to a single machine, or each of them can
    use their own machine, or a mix of both.
\end{itemize}
In principle, these three dimensions are independent from each other.
However, we did not observe every possible constellation yet.
We characterize the ones we did see as follows (refer to
\cref{tab:mode-overview} for a summary):
\begin{itemize}
    \item\label{mode:pp}
    \textbf{Pair Programming} (\textbf{PP}):
    Two developers sit next to each other and work on one single machine.

    \item\label{mode:ppao}
    \textbf{Pair Programming with Active Observer} (\textbf{PPao}):
    Like PP, but one partner occasionally looks up things on her own machine.
    The pair still works on one task though.

    \item\label{mode:sbs}
    \textbf{Side-by-Side Programming} (\textbf{SbS}):
    Two developers sit next to each other and work on their own separated
    but related tasks.

    \item\label{mode:mob}
    \textbf{Mob Programming} (\textbf{Mob}):
    Three or more developers sit close to each other (e.g. in a conference room
    or next to each other in a row) and work on anything between one single
    machine, and one machine per developer.
    We have observed only fully co-located Mob settings.

    \item\label{mode:rpp}
    \textbf{Remote Pair Programming} (\textbf{RPP}):
    Two developers are distributed to two different locations and share one
    screen, i.e., there is one developer who ``owns'' the machine, and the other
    developer may have view-only access (screen-sharing, e.g., via Skype) or
    some interaction options (e.g., via TeamViewer).

    \item\label{mode:dpp}
    \textbf{Distributed Pair Programming} (\textbf{DPP}):
    Two developers are distributed over two different locations and they each
    can interact with their development environment independently but their
    actions are synchronized.
    All our DPP pairs used Saros as their tool \cite{SalOezBee10-saros}.%
    \footnote{Project homepage: \url{https://www.saros-project.org}}
\end{itemize}

\begin{table}[h]
    \centering
    \begin{tabular}{P{2.5cm}@{\hspace{5pt}}l@{\hspace{4pt}}ccc}
        \hline
        \multicolumn{1}{l}{\multirow{2}{*}{\textbf{Name}}}
        & \multirow{2}{*}{\textbf{Short}}
        & \multirow{2}{*}{\textbf{\#Dev}}
        & \textbf{Spatial}
        & \textbf{Active}
        \\
        &
        &
        & \textbf{Setting}
        & \textbf{Computers}
        \\
        \hline
        Pair Programming             & PP  & 2  & co-located & 1 \\
        Pair Programming with Active Observer & PPao & 2  & co-located & 1--2 \\
        Side-by-Side Programming     & SbS & 2  & co-located & 2 \\
        Mob Programming              & Mob & 3+ & co-located & 1+ \\
        Remote Pair Programming      & RPP & 2  & distributed & 1 \\
        Distributed Pair Programming & DPP & 2  & distributed & 2 \\
        \hline
    \end{tabular}
    \caption{Collaboration modes in our data collection}
    \label{tab:mode-overview}
\end{table}

\subsection{Structured Developer Information}
\label{sec:developer-info}

For all 57 developers, we provide the following
structured information:

\begin{itemize}
    \item
    \textbf{Gender} (\textbf{Gnd}):
    We did not ask any participant for their gender specifically.
    We list them as either ``female'' or ``male'' depending on the
    first name they told us.

    \item
    \textbf{Spoken Languages} (\textbf{Lang}):
    We list the developers' spoken languages in decreasing proficiency,
    using ISO 639-1 codes (e.g., ``EN'' for English, ``DE'' for German).
    We only list languages that are actually spoken in our data;
    a dash ``--'' is used as a placeholder, e.g., when it is clear that
    the session language is not the developer's first language and she never
    uses that first language.

    \item
    \textbf{Software Development Experience} (\textbf{Dev.}):
    In some companies, we asked the developers to self-report their
    professional software development experience in years and months
    in the questionnaires (see \cref{fig:questionnaires}).
    In other cases, we retrospectively searched for public profiles on
    professional network sites.

    \textbf{Pair Programming Experience} (\textbf{PP}):
    Here, we only relied on self-reported data from the questionnaires.

    \textbf{Time with Company} (\textbf{Comp.}):
    This aspect we did not ask for in any questionnaire.
    The according numbers stem exclusively from the participants'
    public profiles on career sites.

    For all the experience-related information we use the following conventions:
    A ``\NA'' in a table cell means that no information was ever collected.
    A blank cell means that the developer did fill out the according
    questionnaire, but left the field blank.
    Normal numbers are represented as such, while special markings
    and comments are given in quotation marks, e.g. ``ca.\ 6?'' or
    ``first time''.
    If the same developer was asked the same question on multiple occasions and
    gave different answers, the respective values are separated by a slash
    ``\,/\,''.

    \item
    \textbf{Number of Sessions} (\textbf{\#S}):
    The number of session recordings in our repository the developer is part of.

    \item
    \textbf{Number of Pairs} (\textbf{\#P}):
    The number of distinct pair (or group) constellations in which the developer
    was recorded.
\end{itemize}

\subsection{Structured Session Information}
\label{sec:session-info}

For all 67 sessions, we provide the following structured
information:

\begin{itemize}
    \item
    \textbf{Mode}:
    A characterization of how the developers work together
    (see Section~\ref{sec:modes}).
    We list the predominant mode of the session.
    See the tables' respective captions and \cref{sec:repo}
    for more details on some corner cases.

    \item \textbf{Developers}:
    An incidence matrix of the developers participating in the session.

    \item \textbf{Start}:
    Date and time when the actual session (not just the video) started.

    \item \textbf{Duration} (\textbf{Dur.}):
    Gross session length (from start to dissolving the session),
    including any short breaks.
    An intermittent stand-up meeting of 15 minutes or less counts as short
    whereas a lunch break does not: We then consider the parts
    to be individual sessions even if they where recorded in one sitting.

    \item \textbf{Spoken language(s)} (\textbf{SL}):
    The natural languages used by the participants in the session,
    using ISO 639-1 codes (e.g., ``EN'' for English, ``DE'' for German).

    \item \textbf{Programming Language(s)} (\textbf{PL}):
    The structured languages of the source code files/snippets the developers
    read and/or write in their session.

    \item \textbf{Pre-Session Pair Assessment} (\textbf{Pre}):
    Answer to \emph{task \& pair item} 8: `How well attuned to your partner are you?'
    expressed in German school grades from 1 (best) to 6 (worst).

    \textbf{Post-Session Assessment} (\textbf{Post}):
    Answer to \emph{process item} 1: `How useful was it to solve the task as a pair?'
    expressed in German school grades from 1 (best) to 6 (worst).

    For multiple sessions recorded in one sitting, only the first has Pre data
    and only the last has Post data.
    Pre and Post columns are omitted for companies in
    which no questionnaires were used.
\end{itemize}

\section{Overview of Sessions}
\label{sec:overview}

Table~\ref{tab:session-overview} summarizes the contents of the repository
per company.
In addition to the types and lengths of the recorded sessions,
we also list the quality of the auxiliary data sources, i.e.,
pre- and post-session questionnaires and recordings of the reflective
interviews.

\begin{figure}
\begin{knitrout}
\definecolor{shadecolor}{rgb}{0.969, 0.969, 0.969}\color{fgcolor}
\includegraphics[width=\maxwidth]{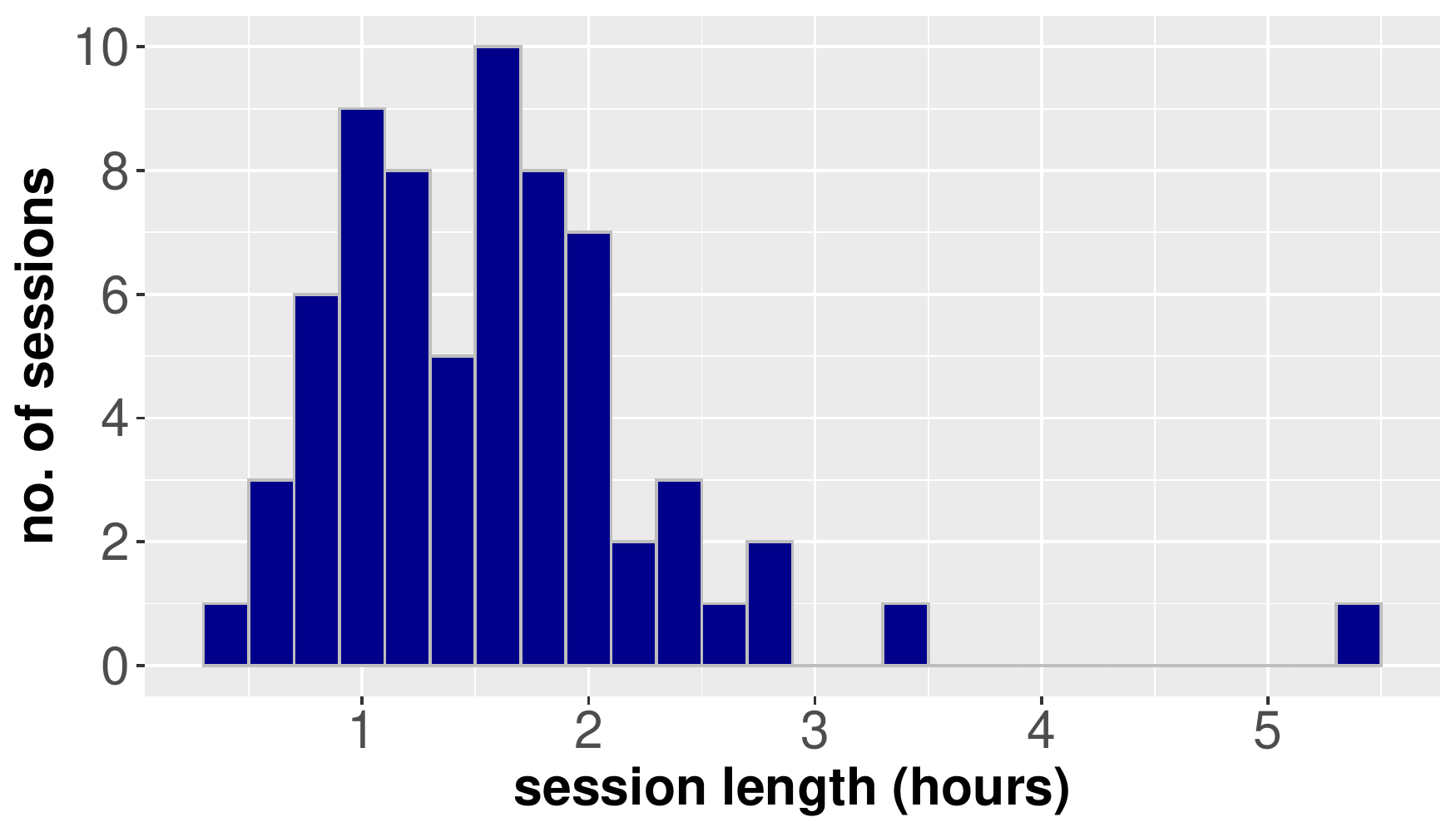} 

\end{knitrout}
\caption{Distribution of Session Lengths}
\label{fig:session-lengths}
\end{figure}

Figure~\ref{fig:session-lengths} is a histogram of the session
lengths.
The vast majority of our sessions runs between 0:45 and 2:30 hours.

Our data set \cite{PPind-meta} contains per-company R files
with all the structured information in the form described in
\cref{sec:developer-info,sec:session-info}.

\begin{table*}[h]
\centering
\begin{tabular}{c|ccc|rrrrrrr|rr|cc|c}
\hline
 \multirow{3}{*}{\textbf{Company}} &
 \multicolumn{12}{c|}{\textbf{Sessions}} &
 \multicolumn{3}{c}{\textbf{Auxiliary Data}}
 \\
 & %
 \multicolumn{3}{c|}{\textbf{Duration}} &
 \multicolumn{6}{c}{\textbf{Mode}} &
 \multirow{2}{*}{$\Sigma$} & 
 \multirow{2}{*}{\textbf{Devs}} &
 \multirow{2}{*}{\textbf{Pairs}} &
 \multicolumn{2}{c|}{\textbf{Questionnaire}} &
 \textbf{Reflection} \\
 & min & avg & max & PP & PPao & SbS & Mob & RPP & DPP
 & %
 & %
 & %
 & Pre
 & Post
 & \textbf{Interview}
 \\
\hline
\textsf{A}&2:22&2:22&2:22&1&&&&&&1&2&1&\yes&\yes&--\\\textsf{B}&1:21&1:37&1:51&4&&&&&&4&2&1&\yes&\yes&--\\\textsf{C}&1:12&1:28&2:10&6&&&&&&6&8&6&\yes&\yes&\yesbut\\\textsf{D}&0:31&1:33&2:23&6&&&&&&6&8&5&\yesbut&\yes&\yesbut\\\textsf{E}&1:17&1:47&2:46&7&&&&&&7&8&6&\yes&\yes&\yesbut\\\textsf{F}&1:45&2:03&2:39&4&&&&&&4&6&4&\yes&\yes&\yesbut\\\textsf{J}&0:42&1:53&5:27&&&&&&9&9&2&1&--&\yesbut&--\\\textsf{K}&0:53&1:40&2:53&8&&&&&&8&4&3&--&\yesbut&\yesbut\\\textsf{L}&0:47&0:53&1:00&&&&&1&1&2&3&2&--&--&--\\\textsf{M}&0:25&0:25&0:25&1&&&&&&1&2&1&--&--&--\\\textsf{N}&0:41&1:35&3:29&&&4&1&&&5&4&3&--&--&\yesbut\\\textsf{O}&0:47&1:13&1:44&2&2&&4&2&&10&5&6&--&--&\yesbut\\\textsf{P}&0:58&1:25&1:42&4&&&&&&4&3&2&--&--&\yes\\
\hline
$\Sigma$&0:25&1:35&5:27&43&2&4&5&3&10&67&57&41\\
\cline{1-13}
\end{tabular}
\caption{Overview of the PP-ind repository of professional pair programming
session recordings including the minimum, average, and maximum durations of
each company's recordings as well as the recording count per collaboration mode
(zeros are omitted for readability).
The modes are explained in \cref{sec:modes}.
Auxiliary data may be \yes--complete, \yesbut--partial, or non-existing.
Questionnaires are \emph{complete} if they were handed out for all sessions
then filled out by both partners individually;
\emph{complete} reflection interviews are all audio-recorded while
\emph{partial} means missing records or hand-written notes only.}
\label{tab:session-overview}
\end{table*}

\section{The Repository}
\label{sec:repo}

In this Section, we characterize each of the
13 companies
and provide summaries for many of the individual sessions.
Characterizing such sessions on a content-level in a uniform
manner is difficult and is beyond the purpose of the current report.
Hence, we only provide summaries in a non-uniform manner and
only for the sessions where we have one at hand from the way
we used the session in our PP research.

We do not generally transcribe our sessions, but provide partial
transcripts (and partial translations) as a growing data set
\cite{PPind-meta}.

\subsection{Company A}
\label{company:A}

Company \textsf{A} develops a web-based content management system (CMS)
in Java and Objective-C.

We recorded a single session (\sRef{AA1}) with developers A1 and A2 in this
company.
Both developers know their domain well and are generally experienced developers.
Developer A1 has very good structural knowledge of the Java frontend and its
individual classes, as well as practical knowledge of the Eclipse IDE and the
Java programming language.
His colleague A2 is more familiar with the backend and the SQL
database, the VIM editor and the UNIX shell, and the Objective-C programming
language.
However, each of them would also be able to work in the other part
of the system.

See Table~\ref{tab:developer-overview-a} for general information on the
developers and Table~\ref{tab:session-overview-a} for structured information on
the recorded session.

\begin{table}[h]
\centering
\begin{tabular}{l|ll|r@{\hspace{2pt}}r|r@{\hspace{2pt}}r|r@{\hspace{2pt}}r|r@{\hspace{2pt}}r}
\hline
\multirow{2}{*}{\textbf{ID}} &
\multirow{2}{*}{\textbf{Gnd}} &
\multirow{2}{*}{\textbf{Lang}} &
\multicolumn{2}{c|}{\textbf{Dev.}} &
\multicolumn{2}{c|}{\textbf{PP}} &
\multicolumn{2}{c|}{\textbf{Comp.}} &
\multirow{2}{*}{\textbf{\#S}} &
\multirow{2}{*}{\textbf{\#P}} \\
& & &
\multicolumn{1}{c}{\textbf{Y}} & \multicolumn{1}{c|}{\textbf{M}} &
\multicolumn{1}{c}{\textbf{Y}} & \multicolumn{1}{c|}{\textbf{M}} &
\multicolumn{1}{c}{\textbf{Y}} & \multicolumn{1}{c|}{\textbf{M}} &
& \\
\hline
A1 & male & DE & 10 & 3 & 2 &  & 4 & 3 & 1 & 1 \\
A2 & male & DE & 7 & 5 & 2 &  & 7 & 5 & 1 & 1 \\

\hline \multicolumn{10}{r}{\textbf{Pair constellations}} & \textbf{%
    1%
} \\
\end{tabular}
\caption{Overview of the \textsf{A} developers.
Experience data as of 2007-01.
See Section~\ref{sec:developer-info} for information on the data and its representation.}
\label{tab:developer-overview-a}
\end{table}

\subsubsection{Session AA1}
\label{session:AA1}

The pair works on fixing inconsistencies across different list views.
They work both in the frontend and the backend code.
Although they know their code base well, they still spend time
understanding peculiarities.
At the end of the session they worked through five different lists and made
them consistent, but learned that one aspect of the business model cannot be
implemented in the backend alone.
They finish with an architectural discussion.

\subsection{Company B}
\label{company:B}

Company \textsf{B} develops a social media platform in PHP and JavaScript.
We visited the company two times to record a total of four PP sessions.

All four sessions are with the same pair of full-stack developers
B1 and B2.
See also Salinger's description of the data collection \cite[pp.~95--99]{Salinger13}.
See Table~\ref{tab:developer-overview-b} for general information on the
developers and Table~\ref{tab:session-overview-b} for structured
information on the recorded sessions.

\emph{Note on Naming Scheme:}
We originally considered the data from the first recording sitting to be
corrupt but were eventually able to mostly recover them.
This is why the earlier sessions are labeled \sRef{BB1} to \sRef{BB3} even
though they take place five months prior to session \sRef{BA1}.

\begin{table}[h]
\centering
\begin{tabular}{l|ll|r@{\hspace{2pt}}r|r@{\hspace{2pt}}r|r@{\hspace{2pt}}r|r@{\hspace{2pt}}r}
\hline
\multirow{2}{*}{\textbf{ID}} &
\multirow{2}{*}{\textbf{Gnd}} &
\multirow{2}{*}{\textbf{Lang}} &
\multicolumn{2}{c|}{\textbf{Dev.}} &
\multicolumn{2}{c|}{\textbf{PP}} &
\multicolumn{2}{c|}{\textbf{Comp.}} &
\multirow{2}{*}{\textbf{\#S}} &
\multirow{2}{*}{\textbf{\#P}} \\
& & &
\multicolumn{1}{c}{\textbf{Y}} & \multicolumn{1}{c|}{\textbf{M}} &
\multicolumn{1}{c}{\textbf{Y}} & \multicolumn{1}{c|}{\textbf{M}} &
\multicolumn{1}{c}{\textbf{Y}} & \multicolumn{1}{c|}{\textbf{M}} &
& \\
\hline
B1 & male & DE & 8 & 10 & 0 & 5 & 0 & 7 & 4 & 1 \\
B2 & male & DE & \multicolumn{2}{c|}{\NA} & 0 & 6 & \multicolumn{2}{c|}{\NA} & 4 & 1 \\

\hline \multicolumn{10}{r}{\textbf{Pair constellations}} & \textbf{%
    1%
} \\
\end{tabular}
\caption{Overview of the \textsf{B} developers.
Experience data as of 2007-04.
See Section~\ref{sec:developer-info} for
information on the data and its representation.}
\label{tab:developer-overview-b}
\end{table}

\subsubsection{Sessions BB1, BB2, and BB3}
\label{session:BB1}
\label{session:BB2}
\label{session:BB3}

The pair implements a new feature from scratch,
going through their complete web development stack:
starting with template and internationalization in session \sRef{BB1},
continuing with controller, model, database layer, and template optics in
session \sRef{BB2}, and
concluding with making their view more interactive through JavaScript
in session \sRef{BB3}.

\textit{Note on Data:}
The webcam and audio recording is jumbled for the last 30 minutes of session
\sRef{BB3}:
While the screen video is continuous, about 10 minutes worth of webcam and
audio are randomly missing, making the other 20 minutes very difficult to
understand due to lack of continuity.

\subsubsection{Session BA1}
\label{session:BA1}

The pair takes over some code of unknown quality written by
outsourced developers.
Technically, they want to implement part of a cache.
In particular, their logic should tell whether the requested data has changed
since a given timestamp.
In their session, they need to understand all existing code for that
functionality (a few dozen lines of PHP code), make some additions, and
encounter difficulties in specifying what exactly their cache should do.
In the first minutes they also struggle with the workstation which is not
theirs and not fully configured to their needs.

\subsection{Company C}
\label{company:C}

Company \textsf{C} develops a geographic information system desktop GUI
application written in Java.
The design of this software uses abstraction elaborately.

Over the course of one week, we recorded 6
sessions involving 8 developers in this company.
See also Plonka's description of the data collection \cite[pp.~64--67]{Plonka12},
and the (German) handout produced for company \textsf{C}
in our data set \cite{PPind-meta}.
See Table~\ref{tab:developer-overview-c} for general information on the
developers and Table~\ref{tab:session-overview-c} for structured
information on the recorded sessions.

\begin{table}[h]
\centering
\begin{tabular}{
    @{\hspace{3pt}}l@{\hspace{3pt}}|
    @{\hspace{3pt}}l@{\hspace{2pt}}l@{\hspace{3pt}}|
    r@{\hspace{2pt}}r|
    r@{\hspace{2pt}}r|
    r@{\hspace{2pt}}r|
    r@{\hspace{2pt}}r
}
\hline
\multirow{2}{*}{\textbf{ID}} &
\multirow{2}{*}{\textbf{Gnd}} &
\multirow{2}{*}{\textbf{Lang}} &
\multicolumn{2}{c|}{\textbf{Dev.}} &
\multicolumn{2}{c|}{\textbf{PP}} &
\multicolumn{2}{c|}{\textbf{Comp.}} &
\multirow{2}{*}{\textbf{\#S}} &
\multirow{2}{*}{\textbf{\#P}} \\
& & &
\multicolumn{1}{c}{\textbf{Y}} & \multicolumn{1}{c|}{\textbf{M}} &
\multicolumn{1}{c}{\textbf{Y}} & \multicolumn{1}{c|}{\textbf{M}} &
\multicolumn{1}{c}{\textbf{Y}} & \multicolumn{1}{c|}{\textbf{M}} &
& \\
\hline
C1 & male & DE & 4 & 2 & 2& & \multicolumn{2}{c|}{\NA} & 1 & 1 \\
C2 & male & DE & 9 & 2 / 3 & ``ca. 6?'' &  & 8 & 5 & 2 & 2 \\
C3 & male & DE & 6 &  & 0 &  & 5 & 9 & 1 & 1 \\
C4 & female & DE & 10 &  & 11 / 12 &  & 6 & 2 & 3 & 3 \\
C5 & male & DE & ``20+'' &  & 20& & 2 & 4 & 1 & 1 \\
C6 & female & DE & 2 & 8 & 6 & 7 & \multicolumn{2}{c|}{\NA} & 1 & 1 \\
C7 & male & DE & 10 &  & 12 &  & 6 & 4 & 2 & 2 \\
C8 & male & DE & 0 & 11 & 3 &  & 1 &  & 1 & 1 \\

\hline \multicolumn{10}{r}{\textbf{Pair constellations}} & \textbf{%
    6%
} \\
\end{tabular}
\caption{Overview of the \textsf{C} developers.
Experience data as of 2008-05.
See Section~\ref{sec:developer-info} for
information on the data and its representation.}
\label{tab:developer-overview-c}
\end{table}

\subsubsection{Session CA1}
\label{session:CA1}

Developers C1 and C2 work on a new form on the system's GUI.
C1 has started working on the form prior to the session;
C2 is new to the task.
In their session, they mostly deal with making their new GUI
component in the form toggleable for which they reuse existing
GUI logic.

\subsubsection{Session CA2}
\label{session:CA2}

Developers C2 and C5 work on a small functional extension. Its
main difficulty lies in understanding and properly applying the
existing design abstractions.
The task was started by C5 prior to the session.
The work consists of design discussion and of moving classes to
another package in the first half of the session and of
implementing and testing a new abstraction in the second half.

\subsubsection{Session CA3}
\label{session:CA3}

Developers C6 and C7 want to implement a new context menu entry which is
only enabled under certain circumstances.
They write test cases for the menu entry to be enabled and disabled, and
refactor code along the way.
This is a simple task, but their IDE freezes for over a minute many
times which slows the pair down a lot.
After 1:20 hours into their session, the pair takes a four-minute break.

\subsubsection{Session CA4}
\label{session:CA4}

Developers C4 and C7 implement a new feature to allow for selection of
multiple graphical features while holding down the \texttt{CTRL} key.
They have to adapt many interfaces in the event handling part of the
software since their feature is the first to react to the \texttt{CTRL}
key.
They write tests, do refactorings, and discuss design a lot.
Much time is lost in the second half of the session due to a
problem with the team's SVN (Subversion) server.
The pair fluently uses two sets of keyboard and mouse.

\subsubsection{Session CA5}
\label{session:CA5}

Developers C3 and C4 start implementing a new feature that allows
users to cut existing geometries (such as points, lines, polygons)
into parts by drawing arbitrary shapes across them.
The pair fluently uses two sets of keyboard and mouse and
has a productive and high-paced session.

\subsubsection{Session CB1}
\label{session:CB1}

Developers C4 and C8 work on a pet project of theirs before official office
hours.
It is written in Java, too, but has nothing to do with the domain of the
\textsf{C} company.

\subsection{Company D}
\label{company:D}
\label{session:DA3}
\label{session:DA4}
\label{session:DA5}
\label{session:DA6}

Company \textsf{D} develops a large customer-relationship system that is based
on Eclipse and comprises about 50 top-level modules written in Java.

During one week, we recorded 6
sessions involving 8 developers in this company.
See also Plonka's description of the data collection \cite[pp.~64--67]{Plonka12}.
See Table~\ref{tab:developer-overview-d} for general information on the
developers and Table~\ref{tab:session-overview-d} for structured
information on the recorded sessions.

\begin{table}[h]
\centering
\begin{tabular}{l|ll|r@{\hspace{2pt}}r|r@{\hspace{2pt}}r|r@{\hspace{2pt}}r|r@{\hspace{2pt}}r}
\hline
\multirow{2}{*}{\textbf{ID}} &
\multirow{2}{*}{\textbf{Gnd}} &
\multirow{2}{*}{\textbf{Lang}} &
\multicolumn{2}{c|}{\textbf{Dev.}} &
\multicolumn{2}{c|}{\textbf{PP}} &
\multicolumn{2}{c|}{\textbf{Comp.}} &
\multirow{2}{*}{\textbf{\#S}} &
\multirow{2}{*}{\textbf{\#P}} \\
& & &
\multicolumn{1}{c}{\textbf{Y}} & \multicolumn{1}{c|}{\textbf{M}} &
\multicolumn{1}{c}{\textbf{Y}} & \multicolumn{1}{c|}{\textbf{M}} &
\multicolumn{1}{c}{\textbf{Y}} & \multicolumn{1}{c|}{\textbf{M}} &
& \\
\hline
D1 & female & DE & 10 &  & 2 &  & 2 & 3 & 1 & 1 \\
D2 & male & DE & 12 &  & 1 & 6 & \multicolumn{2}{c|}{\NA} & 3 & 2 \\
D3 & male & DE &  & 3 &  & 3 &  & 3 & 1 & 1 \\
D4 & male & DE & 1 &  & \multicolumn{2}{c|}{``--''} & 0 & 0 & 2 & 2 \\
D5 & male & DE & 9 &  & 1 & 8 & 1 & 8 & 1 & 1 \\
D6 & male & DE & 11 &  & \multicolumn{2}{c|}{\NA} & 1 & 10 & 1 & 1 \\
D7 & male & --, DE & 4 &  & 2 &  & \multicolumn{2}{c|}{\NA} & 1 & 1 \\
D8 & female & --, DE & 3 & 2 & 3 & 8 & \multicolumn{2}{c|}{\NA} & 2 & 1 \\

\hline \multicolumn{10}{r}{\textbf{Pair constellations}} & \textbf{%
    5%
} \\
\end{tabular}
\caption{Overview of the \textsf{D} developers.
Experience data as of 2008-10.
See Section~\ref{sec:developer-info} for
information on the data and its representation.}
\label{tab:developer-overview-d}
\end{table}

\subsubsection{Session DA1}
\label{session:DA1}

\emph{Note on Data:} No webcam was recorded.

\subsubsection{Session DA2}
\label{session:DA2}

Developer D4 is in his very first week at the company and has never pair
programmed before;
his colleague D3 has been with the company for three months.
It is his first programming job for which he started learning Java.

In their session, they try to implement a new toolbar for one of the
system's many modules.
After some failed attemptes and two long discussions with
additional developers (first D7, then D6), they perform a technically
simple refactoring task which spans many different modules and takes
the rest of session.
Throughout the session, D4 provides D3 with information on programming styles,
technologies, and so on, whereas D3 is more knowledgeable about the code base
and the organizational background.

Underway, the team suspends the session for 15 minutes to participate in
their team's daily stand-up meeting.

\subsection{Company E}
\label{company:E}
\label{session:EA2}
\label{session:EA3}
\label{session:EA4}
\label{session:EA5}
\label{session:EA6}
\label{session:EA7}

Company \textsf{E} develops a graphical desktop application for different
logistics-related tasks in mostly C++ with some
parts written in C\# and Java.

During one week, we recorded 7
sessions involving 8 developers in this company.
See also Plonka's description of the data collection \cite[pp.~64--67]{Plonka12}.
See Table~\ref{tab:developer-overview-e} for general information on the
developers and Table~\ref{tab:session-overview-e} for structured
information on the recorded sessions.

\begin{table}[h]
\centering
\begin{tabular}{l|ll|r@{\hspace{2pt}}r|r@{\hspace{2pt}}r|r@{\hspace{2pt}}r|r@{\hspace{2pt}}r}
\hline
\multirow{2}{*}{\textbf{ID}} &
\multirow{2}{*}{\textbf{Gnd}} &
\multirow{2}{*}{\textbf{Lang}} &
\multicolumn{2}{c|}{\textbf{Dev.}} &
\multicolumn{2}{c|}{\textbf{PP}} &
\multicolumn{2}{c|}{\textbf{Comp.}} &
\multirow{2}{*}{\textbf{\#S}} &
\multirow{2}{*}{\textbf{\#P}} \\
& & &
\multicolumn{1}{c}{\textbf{Y}} & \multicolumn{1}{c|}{\textbf{M}} &
\multicolumn{1}{c}{\textbf{Y}} & \multicolumn{1}{c|}{\textbf{M}} &
\multicolumn{1}{c}{\textbf{Y}} & \multicolumn{1}{c|}{\textbf{M}} &
& \\
\hline
E1 & male & DE & 0 & 5 & 0 & 5 & \multicolumn{2}{c|}{\NA} & 3 & 2 \\
E2 & male & DE & 10 & 8 & 3 / 5 &  & \multicolumn{2}{c|}{\NA} & 3 & 2 \\
E3 & male & DE & 23 & 0 & 0 & 0 & \multicolumn{2}{c|}{\NA} & 2 & 2 \\
E4 & male & DE & 8 & 0 & 0 & 0 & 6 &  & 2 & 2 \\
E5 & male & DE & 10 &  & 2 &  & \multicolumn{2}{c|}{\NA} & 1 & 1 \\
E6 & female & DE & 13 &  & 0 &  & \multicolumn{2}{c|}{\NA} & 1 & 1 \\
E7 & male & DE & 2 &  &  & 2 & \multicolumn{2}{c|}{\NA} & 1 & 1 \\
E8 & male & DE & 2 & 5 & 2 &  & 2 & 7 & 1 & 1 \\

\hline \multicolumn{10}{r}{\textbf{Pair constellations}} & \textbf{%
    6%
} \\
\end{tabular}
\caption{Overview of the \textsf{E} developers.
Experience data as of 2008-10.
See Section~\ref{sec:developer-info} for
information on the data and its representation.}
\label{tab:developer-overview-e}
\end{table}

\subsubsection{Session EA1}
\label{session:EA1}

Prior to the session, developer E2
already tried to debug a display error that leads to routes of ferries being
displayed with an extra segment.
In their PP session, he and colleague E1 go through the unfamiliar source
code step by step with a debugger.
They repeatedly set up a certain state, inspect variables, develop and test
hypotheses, but do not change any code.
They end their session after 80 minutes (without
having made much apparent progress) because of a team meeting.

\emph{Note on Data:}
There are several pauses in the video stream because the
developers accidentally paused and resumed the recording underway
by using the \texttt{F9} and \texttt{F10} keys in their debugger
which the recording software also listened to.

\subsection{Company F}
\label{company:F}
\label{session:FA1}
\label{session:FA2}
\label{session:FA3}
\label{session:FA4}

Company \textsf{F} develops an email marketing tool written in Java.

On three consecutive days, we recorded 4
sessions involving 6 developers in this company.
See also Plonka's description of the data collection \cite[pp.~64--67]{Plonka12}.
See Table~\ref{tab:developer-overview-f} for general information on the
developers and Table~\ref{tab:session-overview-f} for structured
information on the recorded sessions.

\begin{table}[h]
\centering
\begin{tabular}{l|ll|r@{\hspace{2pt}}r|r@{\hspace{2pt}}r|r@{\hspace{2pt}}r|r@{\hspace{2pt}}r}
\hline
\multirow{2}{*}{\textbf{ID}} &
\multirow{2}{*}{\textbf{Gnd}} &
\multirow{2}{*}{\textbf{Lang}} &
\multicolumn{2}{c|}{\textbf{Dev.}} &
\multicolumn{2}{c|}{\textbf{PP}} &
\multicolumn{2}{c|}{\textbf{Comp.}} &
\multirow{2}{*}{\textbf{\#S}} &
\multirow{2}{*}{\textbf{\#P}} \\
& & &
\multicolumn{1}{c}{\textbf{Y}} & \multicolumn{1}{c|}{\textbf{M}} &
\multicolumn{1}{c}{\textbf{Y}} & \multicolumn{1}{c|}{\textbf{M}} &
\multicolumn{1}{c}{\textbf{Y}} & \multicolumn{1}{c|}{\textbf{M}} &
& \\
\hline
F1 & male & DE & 10 &  & 4 &  & 1 & 2 & 1 & 1 \\
F2 & male & DE & 10 &  & 4 &  & 0 & 3 & 2 & 2 \\
F3 & male & DE & 5 &  & 3 &  & 2 & 2 & 2 & 2 \\
F4 & male & DE & 1 & 6 & \multicolumn{2}{c|}{``rarely''} & \multicolumn{2}{c|}{\NA} & 1 & 1 \\
F5 & male & DE & 15 & ``?'' & 5 &  & 7 &  & 1 & 1 \\
F6 & male & DE & 1 & 4 & 1 & 4 & 1 & 4 & 1 & 1 \\

\hline \multicolumn{10}{r}{\textbf{Pair constellations}} & \textbf{%
    4%
} \\
\end{tabular}
\caption{Overview of the \textsf{F} developers.
Experience data as of 2008-11.
See Section~\ref{sec:developer-info} for
information on the data and its representation.}
\label{tab:developer-overview-f}
\end{table}

\emph{Note on Data:}
In all sessions, the pairs worked on a dual-screen setup
but only one screen was recorded.

\subsection{Context J}
\label{company:J}
\label{session:JA3}
\label{session:JA4}
\label{session:JA5}
\label{session:JA6}
\label{session:JA7}
\label{session:JA8}
\label{session:JA9}

Company \textsf{J} is a service provider for public radio broadcasters.
All recorded PP sessions involve the same pair of
developers who do \emph{distributed pair programming} because they
work in different cities.
J2 is employed by \textsf{J} and J1 is a hired consultant.
The pair synchronized their development environments with the
Eclipse plugin ``Saros'',\footnote{%
    Project homepage: \url{https://www.saros-project.org}
}
which allows concurrent editing in all shared files for both
developers.

Their 9 sessions were recorded on four days:
Session \sRef{JA1} on one day, sessions \sRef{JA2} to \sRef{JA9} two weeks later on
three consecutive days.
See also Schenk's description of the data collection \cite[pp.~137--139]{Schenk18}.

J1 and J2 have known each other for a year and
previously shared an office for two months.
Consultant J1 had his first contact with the 
application domain 8 months before the sessions.
See Table~\ref{tab:developer-overview-j} for general information on the
developers and Table~\ref{tab:session-overview-j} for structured
information on the recorded sessions.

\begin{table}[h]
\centering
\begin{tabular}{l|ll|r@{\hspace{2pt}}r|r@{\hspace{2pt}}r|r@{\hspace{2pt}}r|r@{\hspace{2pt}}r}
\hline
\multirow{2}{*}{\textbf{ID}} &
\multirow{2}{*}{\textbf{Gnd}} &
\multirow{2}{*}{\textbf{Lang}} &
\multicolumn{2}{c|}{\textbf{Dev.}} &
\multicolumn{2}{c|}{\textbf{PP}} &
\multicolumn{2}{c|}{\textbf{Comp.}} &
\multirow{2}{*}{\textbf{\#S}} &
\multirow{2}{*}{\textbf{\#P}} \\
& & &
\multicolumn{1}{c}{\textbf{Y}} & \multicolumn{1}{c|}{\textbf{M}} &
\multicolumn{1}{c}{\textbf{Y}} & \multicolumn{1}{c|}{\textbf{M}} &
\multicolumn{1}{c}{\textbf{Y}} & \multicolumn{1}{c|}{\textbf{M}} &
& \\
\hline
J1 & male & DE & 6 & 4 & \multicolumn{2}{c|}{\NA} & 0 & 8 & 9 & 1 \\
J2 & male & DE & 2 & 6 & \multicolumn{2}{c|}{\NA} & 2 & 6 & 9 & 1 \\

\hline \multicolumn{10}{r}{\textbf{Pair constellations}} & \textbf{%
    1%
} \\
\end{tabular}
\caption{Overview of the \textsf{J} developers.
Experience data as of 2013-01.
See Section~\ref{sec:developer-info} for
information on the data and its representation.}
\label{tab:developer-overview-j}
\end{table}

\subsubsection{Session JA1}
\label{session:JA1}

Domain expert J2 had designed and implemented a plugin-based architecture in
Java to monitor and download remote files from the servers of different radio
stations about a year earlier.
J1's role is helping to 
review and clean-up the code together to ease the subsequent
implementation of a new feature.
In the session, they review only one class, try (and fail) to refactor it by
extracting local methods, and ultimately decide to rewrite the whole system
from scratch, which they do two weeks later in sessions \sRef{JA2} to \sRef{JA9}.

\subsubsection{Session JA2}
\label{session:JA2}

The pair starts developing the module from scratch.
In the first part of the session, J2 shows J1 a number of helper
implementations he wrote in the meantime and J1 criticizes them.
Afterwards, they discuss and collect requirements together in a plain
text file.

\subsection{Company K}
\label{company:K}
\label{session:KC3}
\label{session:KC4}

Company \textsf{K} develops and operates a large web portal for many
real estate services using different technologies which are connected
via a microservice architecture.

We collected data on multiple occasions:
Sessions \sRef{KA1} and \sRef{KA2} show an inter-team collaboration,
sessions \sRef{KB1} and \sRef{KB2} show in-team work two months later,
and \sRef{KC1} to \sRef{KC4} take place yet another six months
later after the team had changed its technology stack.
See Table~\ref{tab:developer-overview-k} for general information on the
developers and Table~\ref{tab:session-overview-k} for structured
information on the recorded sessions.

\emph{Note on Naming Scheme:}
Originally, we considered all 8 K-sessions to be from
the same context and numbered them accordingly as ``KA1'' to ``KA8''.
However, we later recognized them as coming from three different contexts 
(different involved systems, even different programming languages).
Therefore, when we mention ``session KA6'' in \cite{ZiePre16-ppknowtrans2},
we actually refer to session \sRef{KC2}.

\begin{table}[h]
\centering
\begin{tabular}{l|ll|r@{\hspace{2pt}}r|r@{\hspace{2pt}}r|r@{\hspace{2pt}}r|r@{\hspace{2pt}}r}
\hline
\multirow{2}{*}{\textbf{ID}} &
\multirow{2}{*}{\textbf{Gnd}} &
\multirow{2}{*}{\textbf{Lang}} &
\multicolumn{2}{c|}{\textbf{Dev.}} &
\multicolumn{2}{c|}{\textbf{PP}} &
\multicolumn{2}{c|}{\textbf{Comp.}} &
\multirow{2}{*}{\textbf{\#S}} &
\multirow{2}{*}{\textbf{\#P}} \\
& & &
\multicolumn{1}{c}{\textbf{Y}} & \multicolumn{1}{c|}{\textbf{M}} &
\multicolumn{1}{c}{\textbf{Y}} & \multicolumn{1}{c|}{\textbf{M}} &
\multicolumn{1}{c}{\textbf{Y}} & \multicolumn{1}{c|}{\textbf{M}} &
& \\
\hline
K1 & male & DE & 0 & 6 & \multicolumn{2}{c|}{\NA} & 0 & 6 & 2 & 1 \\
K2 & male & DE & \multicolumn{2}{c|}{\NA} & 2 & 6 & 0 & 3 & 8 & 3 \\
K3 & male & DE & 8 &  & 2 / 3 &  & 1 &  & 4 & 1 \\
K4 & male & DE & 12 &  & \multicolumn{2}{c|}{\NA} & 1 & 3 & 2 & 1 \\

\hline \multicolumn{10}{r}{\textbf{Pair constellations}} & \textbf{%
    3%
} \\
\end{tabular}
\caption{Overview of the \textsf{K} developers.
Experience data as of 2013-05.
See Section~\ref{sec:developer-info} for
information on the data and its representation.}
\label{tab:developer-overview-k}
\end{table}

\subsubsection{Session KA1 and KA2}
\label{sec:sessions-ka}
\label{session:KA1}
\label{session:KA2}

Developers K1 and K2 come together to work out an API between
their respective teams' subsystems:
K1 is responsible for a mobile app for which K2 writes the endpoint with
Java Spring web framework.\footnote{%
    Project homepage: \url{https://spring.io}
}
Before they can start with their actual task, they first need
to change the target URL of a single link which takes them 45 minutes
and the help of two colleagues, because their development environment was
set up improperly.
Afterwards, K1 explains the data he needs with some dummy JSON
file he prepared and K2 considers which internal microservices are able to
provide these data.

After a lunch break, they create a first implementation in session \sRef{KA2}.
Overall, their pair work involves reading a lot of somewhat-known source code
and existing API specifications, and generates a lot of fresh common ground
between the two.

\subsubsection{Sessions KB1 and KB2}
\label{session:KB1}
\label{session:KB2}

These two sessions were recorded two months after session \sRef{KA1} and \sRef{KA2}.
Developers K2 (who is now more experienced in the domain) and
K3 (who knows more about databases) amend their data model:
First they introduce a new model class and discuss which fields to include.
In the second half they write and debug a database migration to adapt the
database schema.

\subsubsection{Sessions KC1 and KC2}
\label{session:KC1}
\label{session:KC2}

Yet another six months later, developers K2 and K3 work together in their
now changed environment:
The team switched its technology stack from Java to CoffeeScript.\footnote{%
    Project homepage: \url{https://coffeescript.org}
}
The two are in the process of getting to know the jQuery JavaScript
library\footnote{%
    Project homepage: \url{https://jquery.com}
} because they want to write an integration test of an auto-completion
feature, for which they want to programmatically enter characters into an
input field.
In session \sRef{KC1}, they set up their test environment and discuss different
test approaches.
In session \sRef{KC2}, after a lunch break, they try out these approaches
(which do not work as intented) and struggle with the debugger.

\subsection{Context L}
\label{company:L}

L1 is a freelance consultant who offers remote training sessions via
\hyperref[mode:rpp]{RPP} or
\hyperref[mode:dpp]{DPP}.
Individual developers bring their own tasks and work with him remotely.

We recorded two of these training sessions with two different clients L2 and L3.
See Table~\ref{tab:developer-overview-l} for general information on the
developers and Table~\ref{tab:session-overview-l} for structured
information on the recorded sessions.

\begin{table}[h]
\centering
\begin{tabular}{l|ll|r@{\hspace{2pt}}r|r@{\hspace{2pt}}r|r@{\hspace{2pt}}r|r@{\hspace{2pt}}r}
\hline
\multirow{2}{*}{\textbf{ID}} &
\multirow{2}{*}{\textbf{Gnd}} &
\multirow{2}{*}{\textbf{Lang}} &
\multicolumn{2}{c|}{\textbf{Dev.}} &
\multicolumn{2}{c|}{\textbf{PP}} &
\multicolumn{2}{c|}{\textbf{Comp.}} &
\multirow{2}{*}{\textbf{\#S}} &
\multirow{2}{*}{\textbf{\#P}} \\
& & &
\multicolumn{1}{c}{\textbf{Y}} & \multicolumn{1}{c|}{\textbf{M}} &
\multicolumn{1}{c}{\textbf{Y}} & \multicolumn{1}{c|}{\textbf{M}} &
\multicolumn{1}{c}{\textbf{Y}} & \multicolumn{1}{c|}{\textbf{M}} &
& \\
\hline
L1 & male & EN & 11 &  & \multicolumn{2}{c|}{\NA} & 2 & 7 & 2 & 2 \\
L2 & male & EN & 14 &  & \multicolumn{2}{c|}{\NA} & 0 & 1 & 1 & 1 \\
L3 & male & EN & \multicolumn{2}{c|}{\NA} & \multicolumn{2}{c|}{\NA} & \multicolumn{2}{c|}{\NA} & 1 & 1 \\

\hline \multicolumn{10}{r}{\textbf{Pair constellations}} & \textbf{%
    2%
} \\
\end{tabular}
\caption{Overview of the \textsf{L} developers.
Experience data as of 2014-02.
See Section~\ref{sec:developer-info} for
information on the data and its representation.}
\label{tab:developer-overview-l}
\end{table}

\subsubsection{Session LA1}
\label{session:LA1}

Consultant L1 and developer L2 go through a tutorial on the Ruby language.\footnote{%
  Tutorial hompage: \url{http://rubykoans.com}
}
L1 is knowledgeable in Ruby (but still learns some fundamental aspects)
whereas L2 is fairly new to it.

\subsubsection{Session LB1}
\label{session:LB1}

Consultant L1 and developer L3 play around with the impress.js library
for browser-based slideshows.\footnote{%
  Project homepage: \url{https://impress.js.org}
}

\subsection{Company M}
\label{company:M}

Company \textsf{M} develops software for multiple clients in the energy and
logistics sectors.

We recorded a single session with developers M1 and M2.
See Table~\ref{tab:developer-overview-m} for general information on the
developers and Table~\ref{tab:session-overview-m} for structured
information on the recorded sessions.

\begin{table}[h]
\centering
\begin{tabular}{l|ll|r@{\hspace{2pt}}r|r@{\hspace{2pt}}r|r@{\hspace{2pt}}r|r@{\hspace{2pt}}r}
\hline
\multirow{2}{*}{\textbf{ID}} &
\multirow{2}{*}{\textbf{Gnd}} &
\multirow{2}{*}{\textbf{Lang}} &
\multicolumn{2}{c|}{\textbf{Dev.}} &
\multicolumn{2}{c|}{\textbf{PP}} &
\multicolumn{2}{c|}{\textbf{Comp.}} &
\multirow{2}{*}{\textbf{\#S}} &
\multirow{2}{*}{\textbf{\#P}} \\
& & &
\multicolumn{1}{c}{\textbf{Y}} & \multicolumn{1}{c|}{\textbf{M}} &
\multicolumn{1}{c}{\textbf{Y}} & \multicolumn{1}{c|}{\textbf{M}} &
\multicolumn{1}{c}{\textbf{Y}} & \multicolumn{1}{c|}{\textbf{M}} &
& \\
\hline
M1 & male & --, EN & 8 &  & \multicolumn{2}{c|}{\NA} & 2 & 1 & 1 & 1 \\
M2 & male & --, EN & 2 & 2 & \multicolumn{2}{c|}{\NA} & 2 & 2 & 1 & 1 \\

\hline \multicolumn{10}{r}{\textbf{Pair constellations}} & \textbf{%
    1%
} \\
\end{tabular}
\caption{Overview of the \textsf{M} developers.
Experience data as of 2014-10.
See Section~\ref{sec:developer-info} for
information on the data and its representation.}
\label{tab:developer-overview-m}
\end{table}

\subsubsection{Session MA1}
\label{session:MA1}

Developer M2 goes through a number of database tables and asks M1 many
questions about their purpose.
Since M2 has prepared a list of helpful \texttt{SQL SELECT} queries as
a guide, the session is efficient and only lasts 25 minutes.

\subsection{Company N}
\label{company:N}

Company \textsf{N} develops and operates the web platform for a fashion retailer.

We recorded 5 sessions during the company's
``onboarding'' process where a group of new hires is introduced to the
company and its technology stack.
During these sessions, each developer works on his or her own machine to
set up the Docker-\footnote{%
  Project homepage: \url{https://www.docker.com}
}
and AWS-based\footnote{%
  Amazon Web Services: \url{https://aws.amazon.com}
}
development environment.
There is no source code involved, only a multitude of documentation which
the developers try to work through in groups of two or three.
See Table~\ref{tab:developer-overview-n} for general information on the
developers and Table~\ref{tab:session-overview-n} for structured
information on the recorded sessions.

\begin{table}[h]
\centering
\begin{tabular}{l|ll|r@{\hspace{2pt}}r|r@{\hspace{2pt}}r|r@{\hspace{2pt}}r|r@{\hspace{2pt}}r}
\hline
\multirow{2}{*}{\textbf{ID}} &
\multirow{2}{*}{\textbf{Gnd}} &
\multirow{2}{*}{\textbf{Lang}} &
\multicolumn{2}{c|}{\textbf{Dev.}} &
\multicolumn{2}{c|}{\textbf{PP}} &
\multicolumn{2}{c|}{\textbf{Comp.}} &
\multirow{2}{*}{\textbf{\#S}} &
\multirow{2}{*}{\textbf{\#P}} \\
& & &
\multicolumn{1}{c}{\textbf{Y}} & \multicolumn{1}{c|}{\textbf{M}} &
\multicolumn{1}{c}{\textbf{Y}} & \multicolumn{1}{c|}{\textbf{M}} &
\multicolumn{1}{c}{\textbf{Y}} & \multicolumn{1}{c|}{\textbf{M}} &
& \\
\hline
N1 & male & DE & \multicolumn{2}{c|}{\NA} & \multicolumn{2}{c|}{\NA} & 0 & 0 & 2 & 1 \\
N2 & male & DE & \multicolumn{2}{c|}{\NA} & \multicolumn{2}{c|}{\NA} & 0 & 0 & 3 & 2 \\
N3 & female & DE & \multicolumn{2}{c|}{\NA} & \multicolumn{2}{c|}{\NA} & 0 & 0 & 3 & 2 \\
N4 & male & DE & \multicolumn{2}{c|}{\NA} & \multicolumn{2}{c|}{\NA} & 0 & 0 & 3 & 2 \\

\hline \multicolumn{10}{r}{\textbf{Pair/group constellations}} & \textbf{%
    3%
} \\
\end{tabular}
\caption{Overview of the \textsf{N} developers.
Experience data as of 2016-01.
See Section~\ref{sec:developer-info} for
information on the data and its representation.}
\label{tab:developer-overview-n}
\end{table}

\subsubsection{Session NA1 and NA3}
\label{session:NA1}
\label{session:NA3}

Developers N1 and N2 sit next to each other and work through the documentation,
but each configures their own machine.
They take a 1-hour break between \sRef{NA1} and \sRef{NA3}.

\emph{Note on Data:}
While N1's audio, screen, and webcam are recorded, 
only the audio is available for N2.

\subsubsection{Session NA2 and NA4}
\label{session:NA2}
\label{session:NA4}

Analogous to \sRef{NA1} and \sRef{NA3}, including the break and
N4 audio-only limitation.

\subsubsection{Session NA5}
\label{session:NA5}

One week later, developers N2, N3, and N4 sit together at one table, each of
them still trying to set up their respective development environments.
They exchange ideas and insights for three and a half hours.

\emph{Note on Data:}
N4's webcam is not recorded, but all other channels are
(i.e., 3$\times$ audio, 3$\times$ screencast, and 2$\times$ webcam).

\subsection{Company O}

Company \textsf{O} develops a web-based project planning tool using CoffeeScript in
both frontend and backend.

During a four-week observation period we recorded 10
sessions on 4 recording days.
The team employed normal pair programming (\hyperref[mode:pp]{PP}),
mob programming (\hyperref[mode:mob]{Mob}),
side-by-side programming (\hyperref[mode:sbs]{SbS}),
and remote pair programming (\hyperref[mode:rpp]{RPP}).
See Table~\ref{tab:developer-overview-o} for general information on the
developers and Table~\ref{tab:session-overview-o} for structured
information on the recorded sessions.

\begin{table}[h]
\centering
\begin{tabular}{
  l|
  @{\hspace{3pt}}l@{\hspace{3pt}}l@{\hspace{3pt}}|
  r@{\hspace{2pt}}r|
  r@{\hspace{2pt}}r|
  r@{\hspace{2pt}}r|
  r@{\hspace{2pt}}r
}
\hline
\multirow{2}{*}{\textbf{ID}} &
\multirow{2}{*}{\textbf{Gnd}} &
\multirow{2}{*}{\textbf{Lang}} &
\multicolumn{2}{c|}{\textbf{Dev.}} &
\multicolumn{2}{c|}{\textbf{PP}} &
\multicolumn{2}{c|}{\textbf{Comp.}} &
\multirow{2}{*}{\textbf{\#S}} &
\multirow{2}{*}{\textbf{\#P}} \\
& & &
\multicolumn{1}{c}{\textbf{Y}} & \multicolumn{1}{c|}{\textbf{M}} &
\multicolumn{1}{c}{\textbf{Y}} & \multicolumn{1}{c|}{\textbf{M}} &
\multicolumn{1}{c}{\textbf{Y}} & \multicolumn{1}{c|}{\textbf{M}} &
& \\
\hline
O1 & male & DE, EN & 10 & 8 & \multicolumn{2}{c|}{\NA} & 1 &  & 7 & 4 \\
O2 & male & DE, EN & \multicolumn{2}{c|}{\NA} & \multicolumn{2}{c|}{\NA} & 0 & 5 & 2 & 1 \\
O3 & female & --, EN, DE & 1 & 1 & \multicolumn{2}{c|}{\NA} & 0 & 1 & 9 & 5 \\
O4 & male & DE, EN & \multicolumn{2}{c|}{\NA} & \multicolumn{2}{c|}{\NA} & 0 & 5 & 9 & 5 \\
O6 & male & DE, EN & \multicolumn{2}{c|}{\NA} & \multicolumn{2}{c|}{\NA} & \multicolumn{2}{c|}{\NA} & 1 & 1 \\

\hline \multicolumn{10}{r}{\textbf{Pair/group constellations}} & \textbf{%
    6%
} \\
\end{tabular}
\caption{Overview of the \textsf{O} developers.
Experience data as of 2016-06.
See Section~\ref{sec:developer-info} for
information on the data and its representation.}
\label{tab:developer-overview-o}
\end{table}

\subsubsection{Sessions OA1 and OA2}
\label{session:OA1}
\label{session:OA2}

Developers O3 and O4 are tasked with writing a test case for some new functionality.
Even though they get some help from a colleague along the way, they do not make
progress in sessions \sRef{OA1} and \sRef{OA2} (separated by a lunch break).
There are multiple reasons for this:
They neither know that part of the production code nor the underlying
technology (React and Redux\footnote{%
    Project homepages: \url{https://reactjs.org} and \url{https://redux.js.org}
}) nor their development environment so they resort to ``console.log''-debugging
for which they have to rebuild the software in three-minute cycles.
The pair speaks English throughout the session, which is neither developer's
first language.

\emph{Note on Data:}
Session \sRef{OA1} started as plain PP with only one screen.
However, O4 started using his laptop along the way (and continued in \sRef{OA2}),
but his screen is not recorded.
Additionally, there is no webcam recording at all for session \sRef{OA1}.

\subsubsection{Sessions OA3, OA4, and OA5}
\label{session:OA3}
\label{session:OA4}
\label{session:OA5}

Four developers of the team work on a bug that causes some rendering issues
on drag-and-drop actions in a calendar view.
Between \sRef{OA3} and \sRef{OA4} there is a 10 minute break;
between \sRef{OA4} and \sRef{OA5}, there is a 15 minute break, during which developers O2 and
O4 left the group, leaving just O1 and O3 who 
amend test cases, refactor the production code, and eventually fix the bug.
Along the way, O1 explains general software development principles to O3.

\subsubsection{Sessions OA6 and OA7}
\label{session:OA6}
\label{session:OA7}

Developers O3 and O4 try to understand a performance issue in their web application
for about an hour.
Twenty minutes later, O4 continues this task with O1.
Both sessions are done via a Skype call with audio and screensharing but no webcam.

\emph{Note on Data:} No webcam was recorded.

\subsubsection{Sessions OA8, OA9 and OA10}
\label{session:OA8}
\label{session:OA9}
\label{session:OA10}

The three developers O1, O3, and O4 work on a bug in three sessions.
At first, they investigate production code and test code to
understand the reason for a newly failing test case that is
unrelated to the bug:
They changed some implementation but did not
adapt the mock objects used in the tests accordingly.
They eventually adapt the mock objects and write new test cases to reproduce
the bug.
Developer O1 leaves session \sRef{OA8} after 17 minutes, but is again part of the 
group in sessions \sRef{OA9} and \sRef{OA10}.

\subsection{Company P}

Company \textsf{P} develops and operates the web platform for a car part retailer.
The website is consists of multiple large PHP apps.

During a one-week stay, we recorded 4 sessions
with 3 developers.
See Table~\ref{tab:developer-overview-p} for general information on the
developers and Table~\ref{tab:session-overview-p} for structured
information on the recorded sessions.

\begin{table}[h]
\centering
\begin{tabular}{l|ll|r@{\hspace{2pt}}r|r@{\hspace{2pt}}r|r@{\hspace{2pt}}r|r@{\hspace{2pt}}r}
\hline
\multirow{2}{*}{\textbf{ID}} &
\multirow{2}{*}{\textbf{Gnd}} &
\multirow{2}{*}{\textbf{Lang}} &
\multicolumn{2}{c|}{\textbf{Dev.}} &
\multicolumn{2}{c|}{\textbf{PP}} &
\multicolumn{2}{c|}{\textbf{Comp.}} &
\multirow{2}{*}{\textbf{\#S}} &
\multirow{2}{*}{\textbf{\#P}} \\
& & &
\multicolumn{1}{c}{\textbf{Y}} & \multicolumn{1}{c|}{\textbf{M}} &
\multicolumn{1}{c}{\textbf{Y}} & \multicolumn{1}{c|}{\textbf{M}} &
\multicolumn{1}{c}{\textbf{Y}} & \multicolumn{1}{c|}{\textbf{M}} &
& \\
\hline
P1 & male & DE & 5& & 2& & 2& & 4 & 2 \\
P2 & male & DE & 6& & 3 & 6 & 3 & 6 & 2 & 1 \\
P3 & male & DE & 5& & 2& & 2& & 2 & 1 \\

\hline \multicolumn{10}{r}{\textbf{Pair/group constellations}} & \textbf{%
    2%
} \\
\end{tabular}
\caption{Overview of the \textsf{P} developers.
Experience data as of 2018-06.
See Section~\ref{sec:developer-info} for
information on the data and its representation.}
\label{tab:developer-overview-p}
\end{table}

\subsubsection{Sessions PA1 and PA2}
\label{session:PA1}
\label{session:PA2}

In session \sRef{PA1}, developers P1 and P2 review a database
migration written by P1 and discuss the requirements
that led to the database schema change in the first place.
They continue after their lunch break with session \sRef{PA2} where they test
and debug the migration and end up changing test cases to remove embedded
false assumptions.

\subsubsection{Sessions PA3 and PA4}
\label{session:PA3}
\label{session:PA4}

Developers P1 (who is more knowledgeable in the backend) and P3
(more frontend) continue the implementation of a new API
endpoint which P3 already started.
In \sRef{PA3}, P3 shows his existing implementation for which they write
tests; P1 explains backend-related software development best practices.
On the next day, in session \sRef{PA4}, they implement the
database access which causes them problems because of some idiosyncrasy of
their object-relational (OR) mapper.
There is a 13-minute break in session \sRef{PA4} due to a spontaneous team meeting.

\section{Discussion}
\label{sec:discussion}

The data collection procedure described in \cref{sec:protocol}
has a number of properties which affect the recordings' content.

\subsection{Limitation of Scope}

Due to the method's design, our recordings will not reflect all
kinds of relevant PP situations that occur in practice.
Here is a discussion of the likely coverage gaps:
\begin{itemize}
    \item
    \textbf{Not all companies:}
    Due to our naturalistic approach (we did not \emph{request} developers to
    use PP), we did not target companies with little or no pair programming
    usage.

    \item
    \textbf{Not all developers:}
    All recordings are voluntary and some developers may not want to be
    recorded.
    In company~\textsf{P}, for example,
    one team member did use PP,
    but did not want to be part of the data collection.

    \item
    \textbf{No short sessions:}
    The majority of the sessions in the repository is one hour or
    longer, the shortest one is 25 minutes long.
    In discussions with practitioners, however, some reported
    common session lengths of 10 or 15 minutes, often started
    spontaneously and very informally.
    Since our recording setup poses an overhead to the normal work flow,
    such \emph{ad hoc} pairings are difficult to record.
    Conversely, once pairs had gone through setting up
    a recording session, they possibly stayed in it longer than they
    normally would have.

    \item
    \textbf{No tense situations:}
    The mere presence of a researcher on site may be regarded as a distraction.
    In companies \textsf{O} and \textsf{P} there were multiple months
    between the first discussion and the start of the main data collection,
    and in both cases it was due to the Scrum Masters wanting to postpone the
    research activity until a turbulent phase in their respective project was
    over.
    A second data collection phase in company~\textsf{O} did not happen because
    of immense time pressure for the software developers---even though both
    developers and Scrum Masters were very happy with the insights from the
    first round.
    It is not clear whether any pair programming was done in these stressful
    phases.
\end{itemize}
There are other limitations of the data, which are not strictly due to design,
but due to practicalities of getting in contact with a company and traveling.
\begin{itemize}
  \item
    \textbf{Western Cultural Background:}
    All companies are based in Germany with the exception of company 
    \textsf{M} which is in Oslo, Norway.
  \item
    \textbf{Language Limitations:}
    Most developers are native German speakers.
    The \textsf{L}-developers are the only native English speakers
    (but the \textsf{L}-sessions lacked an
    organizational background);
    the \textsf{M}- and \textsf{O}-developers use English as their
    work language.
\end{itemize}

\subsection{Effects of Recording Infrastructure}
\label{sec:recording-effects}

In companies \textsf{A} to \textsf{D}, we have also equipped
the developers' IDEs
with a plugin to collect technical information on their current activities,
focus, etc.\ (see \cite[pp.~85 \& 461--479]{Salinger13}).
This led to some artifacts in the programming sessions.
For instance, in \sRef{CA2}
the developers spent 1:20 minutes trying (and failing) to look up an ID
from the issue tracker on the (remote) development computer before tabbing out
of the RDP session to find the necessary information on the
(local) recording computer within five seconds.
They had first avoided this to ensure continuous data
collection.
In session \sRef{CA3}, the IDE was repeatedly and unexplainedly unresponsive
for 1:20 minutes at a time (totaling about one third(!) of the whole session).
This may have been due to a defect in the data collection plugin.

There were several instances of developers not working on their own machine
and this affecting their work:
In sessions \sRef{CA2} and \sRef{BA1}, the developers spent some time to get comfortable
with their IDE as some options are not set as they were used to;
session \sRef{DA2} starts with several minutes of the pair waiting for an SVN
update to complete since the workspace had not been used for a while.

The recording infrastructure was not always fully compatible with the local
circumstances and the developers' habits;
in session \sRef{EA1}, the same keyboard shortcut had two
meanings that were active at the same time:
Stepping in the IDE's debugger and pausing for our Camtasia recording software.
Not only did this lead to some gaps in the screencast, but appears to have also
confused the developers as the continuous audio recording reveals.
All pairs in company \textsf{F} appear to have used two monitors, but
the recording setup at the time was not able to capture both, so the
screencast is missing one half.

Although wireless microphones and webcam were supposed to not bother the
developers, they occasionally fiddled with them, e.g., before leaving the desk
for a minute and again upon their return.
One pair knocked over the webcam from its tripod;
another pair took a break together to get some candies and wandered beyond the
wireless transmitter radius while still talking about their task.

Reports on how the subjects felt regarding the data collection are available
from the \textsf{C}-developers only, some of which say they felt being watched
while others claim to have forgotten the camera after five minutes.%

\subsection{Effects of Pre-Existing Notions}
\label{sec:preexisting-notions}

We had pre-existing notions of what the social reality of industrial
software development looks like which were deeply embedded in the
data collection process and affect its outcomes.

\subsubsection{Not Recording All Aspects}

The first such notion is \emph{pair programming} itself.
Our research started with the
text-book definition of `two developers jointly working on one computer'.
Neither of these quantities is fixed in everyday industrial development,
but they \emph{are} fixed in the session recording:
Screencast software, microphones, and camera angle are all set up for
\emph{two} developers on \emph{one} computer.
However, developers may suddenly open their own laptop or interact with other
developers who are out of reach of the microphones and/or beyond the camera
angle.

Another impact from the research interest on the way data was collected can be
seen in the sessions recorded by Plonka,
who was initially interested in the \emph{driver/navigator} metaphor.
In order to easily see who is control of keyboard and mouse,
the camera angle of sessions in companies \textsf{C}, \textsf{D},
\textsf{E}, and \textsf{F} centers on the developers' hands
and so occasionally cuts off their faces.
Sessions from the other companies focus on the developers' faces instead.

\subsubsection{Affecting Developer Behavior}

The second, related notion is that a \emph{PP session} pertaining to a
\emph{task} is a meaningful unit of a software developer's workday.
However, some companies form pairs independent of concrete
tasks for multiple days on end during which the pair members behave as
\emph{one}, taking coffee breaks together without really starting or ending a
``session''.
In contrast, our data collection procedure described above is session-centric.
Questionnaires before and/or after the recording frame the session in two
senses.
First, they introduce a ceremonial start and end:
In the beginning of sessions \sRef{CA2} and \sRef{EA1}, the developers filling
out the questionnaire was accidentally recorded---it took them more than nine
minutes, an unnatural intrusion in their work.
Second, the pre-session questionnaire asked the developers to think about the
work time ahead.
In particular, the questionnaire asked for task classification and description,
a characterization of the expected difficulties,
and an estimated time to completion (see \cref{fig:questionnaires}).
Although this may yield valuable context information for the researcher,
it may impose an unnatural focus on the developers by making them think about
aspects they would not have thought about had it not been for the recording.

Another effect of session-centrism can be observed in multiple session
recordings in various companies:
Pair members are occasionally interrupted by their colleagues with technical or
organizational concerns.
A common and unnatural reaction of the pairs in the recordings is
to send away the interrupter unsatisfied
as if to protect the integrity of the data
collection.

The recordings in company~\textsf{E} are peculiar in another way,
which also possibly indicates an intention of the developers to protect the data
collection:
The work station for the session recordings was set up in a meeting room,
so that the pairs were secluded from the rest of their team.

Such effects are more pronounced in the recordings that were done under a
``Understand PP'' headline (see \cref{tab:context}).
For instance, the pairings in companies \textsf{C} to \textsf{F} do not appear
to be holding up to the naturalistic ideal as only \emph{one} of the
overall 21 pairs was recorded twice.\footnote{%
    The pair E1/E2 worked together in \sRef{EA1} and \sRef{EA6}.
    Sessions \sRef{DA5} and \sRef{DA6}
    also feature the same pair (D2/D8),
    but were recorded in one sitting.
}
In the spirit of a ``workshop'', the developers could write their names
on a list with a morning and an afternoon slot if they wanted to be recorded.

In company~\textsf{P}, the team at one point
discussed how the next pair should be formed based on what we at the time
understood as organizational constraints.
However, in the next reflective interview, the developers revealed that
they intended to give another, previously not-recorded colleague the chance to
also benefit from the feedback we would provide.

\subsection{Summary of Data Quality}
\label{sec:data-quality-summary}

Notwithstanding the above limitations of the data collection,
the session repository comprises diverse, realistic, detailed data.
At the time of writing, it contains
67 recordings from
13 different companies featuring
57 different professional software developers who
worked together in
41
different constellations of (mostly) two members.

In these sessions, the developers worked on actual industrial tasks for as
long as they wanted,
and in most cases also freely chose who to work with and when
to start.
The exception here are the
23
sessions from companies \textsf{C} to \textsf{E},
where the developers had to sign up for either a morning of an afternoon slot,
and were possibly inclined to work with partners they would normally not pair
with in the prospect of learning something in the reflective interview.
It can also not be ruled out that the developers in all companies
worked in pairs more often for our recordings than normal.

\begin{figure*}
    \includegraphics{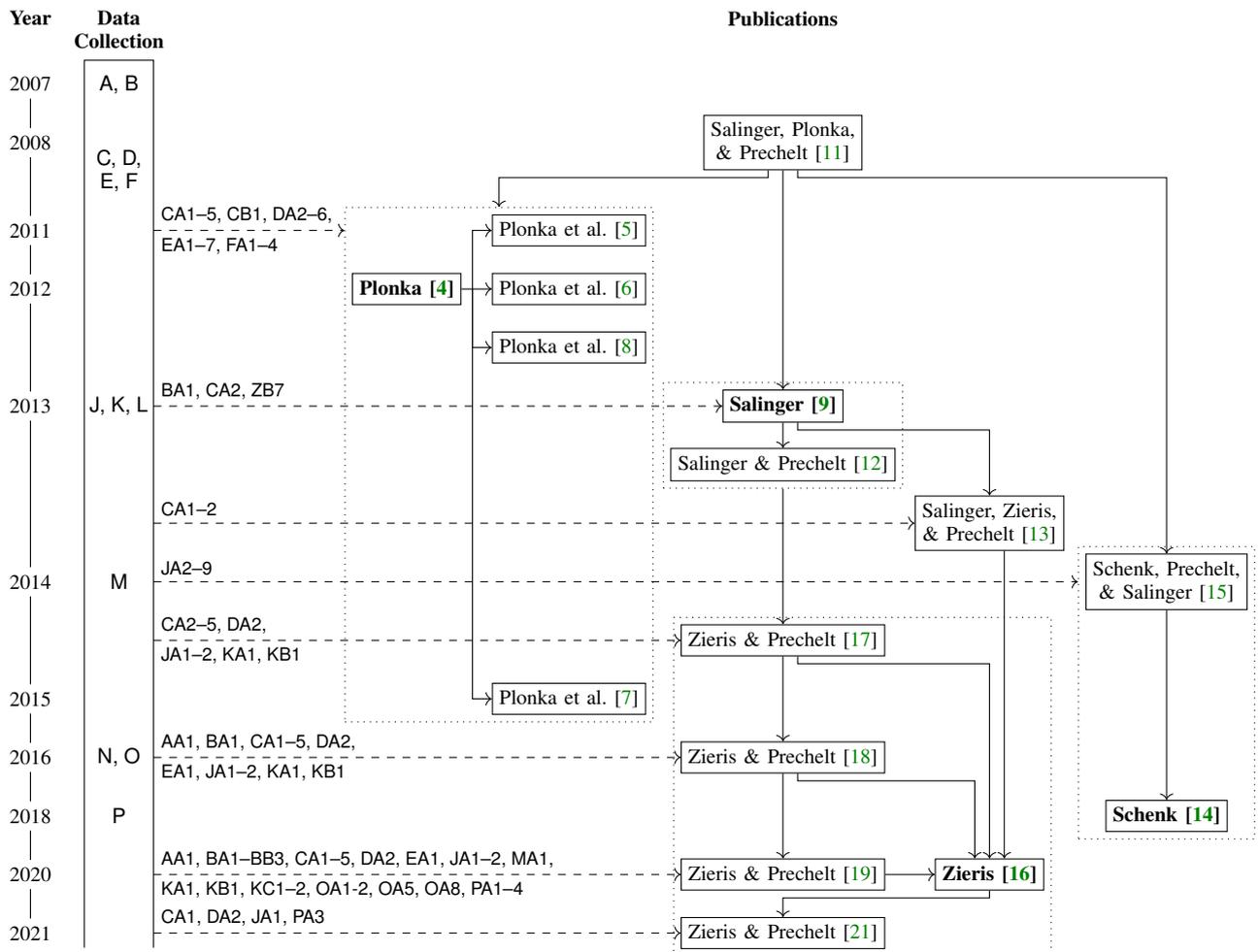}
    \captionsetup{width=\linewidth}
    \caption[]{Timeline of data collection and scientific publications originating in
        our research group.
        Arrows between publications
        \tikz{\draw[->] (0,0) -- (3ex,0);}
        indicate reuse of ideas or building on results;
        arrows from data collection
        \tikz{\draw[->, dashed] (0,0) -- (3ex,0);}
        indicate which PP sessions were analyzed.
        PhD theses are set \textbf{bold}.
        (Note: \textsf{ZB7} is a non-industrial PP session with students.)}
    \label{fig:agse-research-full}
\end{figure*}

All of the above concerns may affect the \emph{frequency} of phenomena
(such as more or fewer conflict situations, more or less easy tasks, or more or
less fatigue due to longer sessions), but none of these appear
likely to produce entirely artificial behavior.
Depending on the make up of concrete studies (e.g., qualitative or quantitative
approach), the above considerations need to be kept in mind.

\section{Usage in Publications}
\label{sec:publications}

See Figure~\ref{fig:agse-research-full} for an overview of all publications
from our research group that build on the data of the PP-ind repository
described in this document.

\begin{table*}[h]
\centering
\begin{tabular}{c|l|@{\hspace{5pt}}c@{\hspace{5pt}}c@{\hspace{5pt}}|clll|c@{\hspace{4pt}}cc@{\hspace{4pt}}c}\hline \multirow{2}{*}{\textbf{ID}}& \multirow{2}{*}{\textbf{Mode}}& \multicolumn{ 2 }{c|}{\textbf{Developer}}& \multirow{2}{*}{\textbf{Start}}& \multirow{2}{*}{\textbf{Dur.}}& \multirow{2}{*}{\textbf{SL}}& \multirow{2}{*}{\textbf{PL}}& \multicolumn{2}{c}{\multirow{2}{*}{\textbf{Pre}}}& \multicolumn{2}{c}{\multirow{2}{*}{\textbf{Post}}}\\
& &  A1 & A2 & & & & & & & & \\\hline
\sRef{AA1} & PP & X & X & 2007-01-26 13:43 & 2:22 & DE & Java, Objective-C, SQL, HTML, Tcl & 2 & 1 & 2 & 1 \\

\hline
\end{tabular}
\caption{Overview of the \textsf{A} sessions. See Section~\ref{sec:session-info} for
information on the data and its representation.}
\label{tab:session-overview-a}
\end{table*}

\begin{table*}[h]
\centering
\begin{tabular}{c|l|@{\hspace{5pt}}c@{\hspace{5pt}}c@{\hspace{5pt}}|clll|c@{\hspace{4pt}}cc@{\hspace{4pt}}c}\hline \multirow{2}{*}{\textbf{ID}}& \multirow{2}{*}{\textbf{Mode}}& \multicolumn{ 2 }{c|}{\textbf{Developer}}& \multirow{2}{*}{\textbf{Start}}& \multirow{2}{*}{\textbf{Dur.}}& \multirow{2}{*}{\textbf{SL}}& \multirow{2}{*}{\textbf{PL}}& \multicolumn{2}{c}{\multirow{2}{*}{\textbf{Pre}}}& \multicolumn{2}{c}{\multirow{2}{*}{\textbf{Post}}}\\
& &  B1 & B2 & & & & & & & & \\\hline
\sRef{BB1} & PP & X & X & 2007-04-27 13:25 & 1:21 & DE & PHP, HTML & 2 & 2 & \multicolumn{2}{c}{\NA} \\
\sRef{BB2} & PP & X & X & 2007-04-27 16:51 & 1:51 & DE & PHP, HTML, SQL, JavaScript, CSS & \multicolumn{2}{c}{\NA} & \multicolumn{2}{c}{\NA} \\
\sRef{BB3} & PP & X & X & 2007-04-27 18:58 & 1:32 & DE & PHP, HTML, JavaScript & \multicolumn{2}{c}{\NA} & \multicolumn{2}{c}{\NA} \\
\sRef{BA1} & PP & X & X & 2007-09-14 13:38 & 1:47 & DE & PHP & 1- & 2 & 2 & 2 \\

\hline
\end{tabular}
\caption{Overview of the \textsf{B} sessions. See Section~\ref{sec:session-info} for
information on the data and its representation.}
\label{tab:session-overview-b}
\end{table*}

\begin{table*}[h]
\centering
\begin{tabular}{c|l|@{\hspace{5pt}}c@{\hspace{5pt}}c@{\hspace{5pt}}c@{\hspace{5pt}}c@{\hspace{5pt}}c@{\hspace{5pt}}c@{\hspace{5pt}}c@{\hspace{5pt}}c@{\hspace{5pt}}|clll|c@{\hspace{4pt}}cc@{\hspace{4pt}}c}\hline \multirow{2}{*}{\textbf{ID}}& \multirow{2}{*}{\textbf{Mode}}& \multicolumn{ 8 }{c|}{\textbf{Developer}}& \multirow{2}{*}{\textbf{Start}}& \multirow{2}{*}{\textbf{Dur.}}& \multirow{2}{*}{\textbf{SL}}& \multirow{2}{*}{\textbf{PL}}& \multicolumn{2}{c}{\multirow{2}{*}{\textbf{Pre}}}& \multicolumn{2}{c}{\multirow{2}{*}{\textbf{Post}}}\\
& &  C1 & C2 & C3 & C4 & C5 & C6 & C7 & C8 & & & & & & & & \\\hline
\sRef{CA1} & PP & X & X &  &  &  &  &  &  & 2008-05-05 13:27 & 1:18 & DE & Java & 3 & 3 & 2 & 2 \\
\sRef{CB1} & PP &  &  &  & X &  &  &  & X & 2008-05-06 07:54 & 1:12 & DE & Java & 2 & 1 & 2 & 2 \\
\sRef{CA2} & PP &  & X &  &  & X &  &  &  & 2008-05-07 11:46 & 1:14 & DE & Java & 3 & 4 & 3 & 1 \\
\sRef{CA3} & PP &  &  &  &  &  & X & X &  & 2008-05-07 15:34 & 2:10 & DE & Java & 2 & 2 & 1 & 1 \\
\sRef{CA4} & PP &  &  &  & X &  &  & X &  & 2008-05-08 10:25 & 1:34 & DE & Java & 1 & 1 & 2 & 2 \\
\sRef{CA5} & PP &  &  & X & X &  &  &  &  & 2008-05-09 10:32 & 1:23 & DE & Java & 4 & 4 & 2 & 3+ \\

\hline
\end{tabular}
\caption{Overview of the \textsf{C} sessions. See Section~\ref{sec:session-info} for
information on the data and its representation.}
\label{tab:session-overview-c}
\end{table*}

\begin{table*}[h]
\centering
\begin{tabular}{c|l|@{\hspace{5pt}}c@{\hspace{5pt}}c@{\hspace{5pt}}c@{\hspace{5pt}}c@{\hspace{5pt}}c@{\hspace{5pt}}c@{\hspace{5pt}}c@{\hspace{5pt}}c@{\hspace{5pt}}|clll|c@{\hspace{4pt}}cc@{\hspace{4pt}}c}\hline \multirow{2}{*}{\textbf{ID}}& \multirow{2}{*}{\textbf{Mode}}& \multicolumn{ 8 }{c|}{\textbf{Developer}}& \multirow{2}{*}{\textbf{Start}}& \multirow{2}{*}{\textbf{Dur.}}& \multirow{2}{*}{\textbf{SL}}& \multirow{2}{*}{\textbf{PL}}& \multicolumn{2}{c}{\multirow{2}{*}{\textbf{Pre}}}& \multicolumn{2}{c}{\multirow{2}{*}{\textbf{Post}}}\\
& &  D1 & D2 & D3 & D4 & D5 & D6 & D7 & D8 & & & & & & & & \\\hline
\sRef{DA1} & PP & X & X &  &  &  &  &  &  & 2008-10-06 14:07 & 2:21 & DE & Java, XML & \multicolumn{2}{c}{``3--4''} & 1 & 1 \\
\sRef{DA2} & PP &  &  & X & X &  &  &  &  & 2008-10-08 10:13 & 2:23 & DE & Java & \multicolumn{2}{c}{``first time''} & 2 & 2.5 \\
\sRef{DA3} & PP &  &  &  & X & X &  &  &  & 2008-10-08 14:01 & 1:05 & DE & XML (Spring) & \multicolumn{2}{c}{6} & 2.5 & 5 \\
\sRef{DA4} & PP &  &  &  &  &  & X & X &  & 2008-10-08 16:22 & 2:00 & DE & Java, XML & \multicolumn{2}{c}{3} & 2 & ``not done'' \\
\sRef{DA5} & PP &  & X &  &  &  &  &  & X & 2008-10-09 10:27 & 0:31 & DE & Java & \multicolumn{2}{c}{1} & \multicolumn{2}{c}{\NA} \\
\sRef{DA6} & PP &  & X &  &  &  &  &  & X & 2008-10-09 13:01 & 0:58 & DE & Java & \multicolumn{2}{c}{\NA} & ``1 / 4'' & 1 \\

\hline
\end{tabular}
\caption{Overview of the \textsf{D} sessions. See Section~\ref{sec:session-info} for
information on the data and its representation. All pairs filled out the pre-session questionnaire together.
In session \sRef{DA2}, the pair talks with two other developers (5 minutes with
D7, then 12 minutes with D6).
In session \sRef{DA6}, the pair worked on two somewhat separate tasks which D2
rated separately in the post-session questionnaire.
}
\label{tab:session-overview-d}
\end{table*}

\begin{table*}[h]
\centering
\begin{tabular}{c|l|@{\hspace{5pt}}c@{\hspace{5pt}}c@{\hspace{5pt}}c@{\hspace{5pt}}c@{\hspace{5pt}}c@{\hspace{5pt}}c@{\hspace{5pt}}c@{\hspace{5pt}}c@{\hspace{5pt}}|clll|c@{\hspace{4pt}}cc@{\hspace{4pt}}c}\hline \multirow{2}{*}{\textbf{ID}}& \multirow{2}{*}{\textbf{Mode}}& \multicolumn{ 8 }{c|}{\textbf{Developer}}& \multirow{2}{*}{\textbf{Start}}& \multirow{2}{*}{\textbf{Dur.}}& \multirow{2}{*}{\textbf{SL}}& \multirow{2}{*}{\textbf{PL}}& \multicolumn{2}{c}{\multirow{2}{*}{\textbf{Pre}}}& \multicolumn{2}{c}{\multirow{2}{*}{\textbf{Post}}}\\
& &  E1 & E2 & E3 & E4 & E5 & E6 & E7 & E8 & & & & & & & & \\\hline
\sRef{EA1} & PP & X & X &  &  &  &  &  &  & 2008-10-27 11:29 & 1:17 & DE & C++ & ``2--3'' & 3 & 2 & 3 \\
\sRef{EA2} & PP &  &  & X & X &  &  &  &  & 2008-10-27 13:18 & 2:46 & DE & XML (Maven) & ``first time'' & ``--'' & 1 & 1 \\
\sRef{EA3} & PP & X &  &  &  & X &  &  &  & 2008-10-28 10:43 & 1:25 & DE & C++ & 3 & 3 & 3 & ``3--2'' \\
\sRef{EA4} & PP &  &  &  & X &  & X &  &  & 2008-10-29 09:37 & 1:52 & DE & C\# & ``--'' & ``0'' & 2 & 2 \\
\sRef{EA5} & PP &  &  & X &  &  &  & X &  & 2008-10-29 13:09 & 1:41 & DE & C++, Java & ``none'' & ``5--6'' & 2 & 2 \\
\sRef{EA6} & PP & X & X &  &  &  &  &  &  & 2008-10-30 10:05 & 1:40 & DE & C++ & ``2--3'' & 3 & 2 & 2 \\
\sRef{EA7} & PP &  & X &  &  &  &  &  & X & 2008-10-31 09:25 & 1:50 & DE & C++ & 2- & 2- & 2 & 2+ \\

\hline
\end{tabular}
\caption{Overview of the \textsf{E} sessions. See Section~\ref{sec:session-info} for
information on the data and its representation.}
\label{tab:session-overview-e}
\end{table*}

\begin{table*}[h]
\centering
\begin{tabular}{c|l|@{\hspace{5pt}}c@{\hspace{5pt}}c@{\hspace{5pt}}c@{\hspace{5pt}}c@{\hspace{5pt}}c@{\hspace{5pt}}c@{\hspace{5pt}}|clll|c@{\hspace{4pt}}cc@{\hspace{4pt}}c}\hline \multirow{2}{*}{\textbf{ID}}& \multirow{2}{*}{\textbf{Mode}}& \multicolumn{ 6 }{c|}{\textbf{Developer}}& \multirow{2}{*}{\textbf{Start}}& \multirow{2}{*}{\textbf{Dur.}}& \multirow{2}{*}{\textbf{SL}}& \multirow{2}{*}{\textbf{PL}}& \multicolumn{2}{c}{\multirow{2}{*}{\textbf{Pre}}}& \multicolumn{2}{c}{\multirow{2}{*}{\textbf{Post}}}\\
& &  F1 & F2 & F3 & F4 & F5 & F6 & & & & & & & & \\\hline
\sRef{FA1} & PP & X & X &  &  &  &  & 2008-11-11 15:06 & 1:45 & DE & Java, XML, SQL & 3 & 3 & 2 & 2 \\
\sRef{FA2} & PP &  & X & X &  &  &  & 2008-11-12 11:07 & 1:59 & DE & Java, XML, SQL & 3 & 3 & 1 & 2 \\
\sRef{FA3} & PP &  &  & X & X &  &  & 2008-11-13 11:06 & 2:39 & DE & Java & 2 & 3 & 2 & 1 \\
\sRef{FA4} & PP &  &  &  &  & X & X & 2008-11-13 15:18 & 1:52 & DE & Java & 2 & 2 & 3 & 1 \\

\hline
\end{tabular}
\caption{Overview of the \textsf{F} sessions. See Section~\ref{sec:session-info} for
information on the data and its representation.}
\label{tab:session-overview-f}
\end{table*}

\begin{table*}[h]
\centering
\begin{tabular}{c|l|@{\hspace{5pt}}c@{\hspace{5pt}}c@{\hspace{5pt}}|clll}\hline \multirow{2}{*}{\textbf{ID}}& \multirow{2}{*}{\textbf{Mode}}& \multicolumn{ 2 }{c|}{\textbf{Developer}}& \multirow{2}{*}{\textbf{Start}}& \multirow{2}{*}{\textbf{Dur.}}& \multirow{2}{*}{\textbf{SL}}& \multirow{2}{*}{\textbf{PL}}\\
& &  J1 & J2 & & & &  \\\hline
\sRef{JA1} & DPP & X & X & 2013-01-31 14:05 & 1:07 & DE & Java \\
\sRef{JA2} & DPP & X & X & 2013-02-13 10:51 & 1:15 & DE & Java \\
\sRef{JA3} & DPP & X & X & 2013-02-13 13:17 & 1:53 & DE & Java \\
\sRef{JA4} & DPP & X & X & 2013-02-13 15:26 & 2:01 & DE & Java \\
\sRef{JA5} & DPP & X & X & 2013-02-14 10:33 & 1:36 & DE & Java \\
\sRef{JA6} & DPP & X & X & 2013-02-14 13:13 & 0:42 & DE & Java \\
\sRef{JA7} & DPP & X & X & 2013-02-14 14:54 & 1:56 & DE & Java \\
\sRef{JA8} & DPP & X & X & 2013-02-15 10:54 & 1:06 & DE & Java \\
\sRef{JA9} & DPP & X & X & 2013-02-15 12:59 & 5:27 & DE & Java \\

\hline
\end{tabular}
\caption{Overview of the \textsf{J} sessions. See Section~\ref{sec:session-info} for
information on the data and its representation.}
\label{tab:session-overview-j}
\end{table*}

\begin{table*}[h]
\centering
\begin{tabular}{c|l|@{\hspace{5pt}}c@{\hspace{5pt}}c@{\hspace{5pt}}c@{\hspace{5pt}}c@{\hspace{5pt}}|clll}\hline \multirow{2}{*}{\textbf{ID}}& \multirow{2}{*}{\textbf{Mode}}& \multicolumn{ 4 }{c|}{\textbf{Developer}}& \multirow{2}{*}{\textbf{Start}}& \multirow{2}{*}{\textbf{Dur.}}& \multirow{2}{*}{\textbf{SL}}& \multirow{2}{*}{\textbf{PL}}\\
& &  K1 & K2 & K3 & K4 & & & &  \\\hline
\sRef{KA1} & PP & X & X &  &  & 2013-03-14 10:37 & 2:00 & DE & Java \\
\sRef{KA2} & PP & X & X &  &  & 2013-03-14 13:15 & 2:53 & DE & Java \\
\sRef{KB1} & PP &  & X & X &  & 2013-05-02 13:45 & 0:53 & DE & Java, SQL \\
\sRef{KB2} & PP &  & X & X &  & 2013-05-02 15:26 & 1:36 & DE & Java, SQL \\
\sRef{KC1} & PP &  & X & X &  & 2013-10-29 11:24 & 0:59 & DE & CoffeeScript \\
\sRef{KC2} & PP &  & X & X &  & 2013-10-29 12:59 & 2:01 & DE & CoffeeScript \\
\sRef{KC3} & PP &  & X &  & X & 2013-11-29 11:00 & 0:53 & DE & CoffeeScript \\
\sRef{KC4} & PP &  & X &  & X & 2013-11-29 11:44 & 2:10 & DE & CoffeeScript \\

\hline
\end{tabular}
\caption{Overview of the \textsf{K} sessions. See Section~\ref{sec:session-info} for
information on the data and its representation.
In session \sRef{KA1}, the pair talks with developer K4 for 8 minutes.}
\label{tab:session-overview-k}
\end{table*}

\begin{table*}[h]
\centering
\begin{tabular}{c|l|@{\hspace{5pt}}c@{\hspace{5pt}}c@{\hspace{5pt}}c@{\hspace{5pt}}|clll}\hline \multirow{2}{*}{\textbf{ID}}& \multirow{2}{*}{\textbf{Mode}}& \multicolumn{ 3 }{c|}{\textbf{Developer}}& \multirow{2}{*}{\textbf{Start}}& \multirow{2}{*}{\textbf{Dur.}}& \multirow{2}{*}{\textbf{SL}}& \multirow{2}{*}{\textbf{PL}}\\
& &  L1 & L2 & L3 & & & &  \\\hline
\sRef{LA1} & RPP & X & X &  & 2014-02-27 20:41 & 1:00 & EN & Ruby \\
\sRef{LB1} & DPP & X &  & X & 2014-03-06 16:11 & 0:47 & EN & HTML, CSS \\

\hline
\end{tabular}
\caption{Overview of the \textsf{L} sessions. See Section~\ref{sec:session-info} for
information on the data and its representation.}
\label{tab:session-overview-l}
\end{table*}

\begin{table*}[h]
\centering
\begin{tabular}{c|l|@{\hspace{5pt}}c@{\hspace{5pt}}c@{\hspace{5pt}}|clll}\hline \multirow{2}{*}{\textbf{ID}}& \multirow{2}{*}{\textbf{Mode}}& \multicolumn{ 2 }{c|}{\textbf{Developer}}& \multirow{2}{*}{\textbf{Start}}& \multirow{2}{*}{\textbf{Dur.}}& \multirow{2}{*}{\textbf{SL}}& \multirow{2}{*}{\textbf{PL}}\\
& &  M1 & M2 & & & &  \\\hline
\sRef{MA1} & PP & X & X & 2014-10-16 11:42 & 0:25 & EN & SQL \\

\hline
\end{tabular}
\caption{Overview of the \textsf{M} sessions. See Section~\ref{sec:session-info} for
information on the data and its representation.}
\label{tab:session-overview-m}
\end{table*}

\begin{table*}[h]
\centering
\begin{tabular}{c|l|@{\hspace{5pt}}c@{\hspace{5pt}}c@{\hspace{5pt}}c@{\hspace{5pt}}c@{\hspace{5pt}}|clll}\hline \multirow{2}{*}{\textbf{ID}}& \multirow{2}{*}{\textbf{Mode}}& \multicolumn{ 4 }{c|}{\textbf{Developer}}& \multirow{2}{*}{\textbf{Start}}& \multirow{2}{*}{\textbf{Dur.}}& \multirow{2}{*}{\textbf{SL}}& \multirow{2}{*}{\textbf{PL}}\\
& &  N1 & N2 & N3 & N4 & & & &  \\\hline
\sRef{NA1} & SbS & X & X &  &  & 2016-01-12 13:30 & 0:47 & DE & -- \\
\sRef{NA2} & SbS &  &  & X & X & 2016-01-12 13:37 & 0:41 & DE & -- \\
\sRef{NA3} & SbS & X & X &  &  & 2016-01-12 15:06 & 0:57 & DE & -- \\
\sRef{NA4} & SbS &  &  & X & X & 2016-01-12 15:12 & 2:04 & DE & -- \\
\sRef{NA5} & Mob &  & X & X & X & 2016-01-18 12:46 & 3:29 & DE & -- \\

\hline
\end{tabular}
\caption{Overview of the \textsf{N} sessions. See Section~\ref{sec:session-info} for
information on the data and its representation.}
\label{tab:session-overview-n}
\end{table*}

\begin{table*}[h]
\centering
\begin{tabular}{c|l|@{\hspace{5pt}}c@{\hspace{5pt}}c@{\hspace{5pt}}c@{\hspace{5pt}}c@{\hspace{5pt}}c@{\hspace{5pt}}|clll}\hline \multirow{2}{*}{\textbf{ID}}& \multirow{2}{*}{\textbf{Mode}}& \multicolumn{ 5 }{c|}{\textbf{Developer}}& \multirow{2}{*}{\textbf{Start}}& \multirow{2}{*}{\textbf{Dur.}}& \multirow{2}{*}{\textbf{SL}}& \multirow{2}{*}{\textbf{PL}}\\
& &  O1 & O2 & O3 & O4 & O6 & & & &  \\\hline
\sRef{OA1} & PPao &  &  & X & X &  & 2016-06-01 10:51 & 1:24 & EN & CoffeeScript \\
\sRef{OA2} & PPao &  &  & X & X & X & 2016-06-01 13:27 & 1:32 & EN & CoffeeScript \\
\sRef{OA3} & Mob & X & X & X & X &  & 2016-06-08 14:55 & 0:59 & DE, EN & CoffeeScript \\
\sRef{OA4} & Mob & X & X & X & X &  & 2016-06-08 16:05 & 0:50 & DE, EN & CoffeeScript \\
\sRef{OA5} & PP & X &  & X &  &  & 2016-06-08 17:11 & 1:09 & EN & CoffeeScript \\
\sRef{OA6} & RPP &  &  & X & X &  & 2016-06-09 14:00 & 0:56 & EN & CoffeeScript \\
\sRef{OA7} & RPP & X &  &  & X &  & 2016-06-09 15:17 & 1:39 & DE & CoffeeScript \\
\sRef{OA8} & PP & X &  & X & X &  & 2016-06-15 13:47 & 1:16 & EN & CoffeeScript \\
\sRef{OA9} & Mob & X &  & X & X &  & 2016-06-15 15:16 & 0:47 & EN & CoffeeScript \\
\sRef{OA10} & Mob & X &  & X & X &  & 2016-06-15 16:50 & 1:44 & EN & CoffeeScript \\

\hline
\end{tabular}
\caption{Overview of the \textsf{O} sessions. See Section~\ref{sec:session-info} for
information on the data and its representation.
    Session \sRef{OA2} started as a \hyperref[mode:ppao]{PP (with Active Observer)} session, and developer O6 joined for
    a period of 14 minutes.
    Session \sRef{OA8} started as a \hyperref[mode:mob]{Mob Programming} session, but O1 had to leave for a
    meeting after 17 minutes.}
\label{tab:session-overview-o}
\end{table*}

\begin{table*}[h]
\centering
\begin{tabular}{c|l|@{\hspace{5pt}}c@{\hspace{5pt}}c@{\hspace{5pt}}c@{\hspace{5pt}}|clll}\hline \multirow{2}{*}{\textbf{ID}}& \multirow{2}{*}{\textbf{Mode}}& \multicolumn{ 3 }{c|}{\textbf{Developer}}& \multirow{2}{*}{\textbf{Start}}& \multirow{2}{*}{\textbf{Dur.}}& \multirow{2}{*}{\textbf{SL}}& \multirow{2}{*}{\textbf{PL}}\\
& &  P1 & P2 & P3 & & & &  \\\hline
\sRef{PA1} & PP & X & X &  & 2018-06-05 11:24 & 0:58 & DE & PHP \\
\sRef{PA2} & PP & X & X &  & 2018-06-05 13:35 & 1:30 & DE & PHP \\
\sRef{PA3} & PP & X &  & X & 2018-06-06 12:23 & 1:31 & DE & PHP \\
\sRef{PA4} & PP & X &  & X & 2018-06-07 11:09 & 1:42 & DE & PHP \\

\hline
\end{tabular}
\caption{Overview of the \textsf{P} sessions. See Section~\ref{sec:session-info} for
information on the data and its representation.}
\label{tab:session-overview-p}
\end{table*}

\printbibliography

\clearpage
\twocolumn

\appendices

\section{Recording Technicalities}
\label{sec:technicalities}
 
We used three different generations of recording setups,
the first of which was developed by Laura Plonka and Stephan Salinger,
the other two by Julia Schenk and Franz Zieris.

\definecolor{clrAudio}{RGB}{56,174,56}
\definecolor{clrWebcam}{RGB}{180,32,168}
\definecolor{clrScreen}{RGB}{7,18,37}
\definecolor{clrScreenM}{RGB}{98,110,132}
\definecolor{clrDev}{RGB}{7,137,172}
\definecolor{clrRec}{RGB}{242,10,10}

\tikzstyle{screen} = [
    rectangle, draw, clrScreen, minimum width=19mm, minimum height=15mm,
    fill=clrScreen!20!white, fill opacity=0.5, text opacity=1
]

\tikzstyle{screenM} = [
    rectangle, draw, clrScreenM, minimum width=19mm, minimum height=15mm,
    fill=clrScreenM!20!white, fill opacity=0.4, text opacity=1
]

\tikzstyle{recorder} = [
    rectangle, draw, clrScreen, minimum width=19mm, minimum height=15mm
]

\tikzstyle{recording} = [
    rectangle, draw, thick, clrRec
]

\tikzstyle{webcam} = [
    rectangle, draw, clrWebcam, minimum width=4mm, minimum height=3mm,
    inner xsep=0.5mm
]

\tikzstyle{audioR} = [
    rectangle, draw, clrAudio, minimum width=4mm, minimum height=3mm
]

\tikzstyle{subject} = [
    circle, draw, clrDev, minimum size=7mm,
    fill=clrDev!20!white, fill opacity=0.6
]

\tikzstyle{webcamV} = [
    clrWebcam
]

\tikzstyle{audioC} = [
    clrAudio
]

\begin{figure*}
    \input{recording-tikz}
    \caption[Different recording setups]%
    {Different recording setups.
        Developers \tikz[baseline=-0.6ex]{\node[subject,minimum size=1.9ex]{};}
        sit in front of a screen
        \tikz[baseline=-0.8ex]{%
            \node[screen, minimum width=2mm, minimum height=2ex,
            inner xsep=2pt, inner ysep=0pt]{\scriptsize\textsf{S}};
        }
        with a webcam
        \tikz[baseline=-0.8ex]{%
            \draw[webcamV] (0,0) -- +(left:1ex)
            node[webcam, anchor=east, minimum width=2mm, minimum height=2ex,
            inner xsep=2pt, inner ysep=0pt]{\scriptsize\textsf{W}};
        }
        and some means to record audio
        \tikz[baseline=-0.8ex]{%
            \draw[audioC] (0,0) -- +(left:1ex)
            node[audioR, anchor=east, minimum width=2mm, minimum height=2ex,
            inner xsep=2pt, inner ysep=0pt]{\scriptsize\textsf{A}};
        }.
        Screen contents are transferred
        \tikz[baseline=-0.8ex]{\draw[-{Latex},gray] (0,0) -- +(right:3ex);}
        to another machine
        (together with audio
        \tikz[baseline=-0.8ex]{%
            \draw[audioC,densely dashed] (0,0) -- +(right:3ex);}
        and/or webcam signal
        \tikz[baseline=-0.8ex]{%
            \draw[webcamV,densely dashed] (0,0) -- +(right:3ex);}%
        ),
        where they are recorded
        \tikz[baseline=-0.8ex]{%
            \node[recording, minimum width=4mm, minimum height=2ex,
            inner sep=0pt]{};
        }
        with a screen capture tool.}
    \label{fig:recording-setups}
\end{figure*}

\subsection*{Generation 1}

Two computers are used in this setup:
The development environment of the developers runs on one machine
whose output is mirrored to another computer where the screen content
is recorded with TechSmith Camtasia.\footnote{%
    Product homepage: \url{https://www.techsmith.com/video-editor.html}
}
On Linux machines (company~\textsf{A}),
developers would work locally while the screen content
is transferred to a Windows machine via VNC (Virtual Network Computing) to
be recorded remotely (see \cref{fig:recording-setup-2}).
On Windows machines
(companies \textsf{C}, \textsf{D}, \textsf{E}, \textsf{F}),
the developers would sit on the recording machine and use RDP (Remote Desktop
Protocol) to log on the development machine, such that the recording device
only needs to run Camtasia and the RDP client
(see \cref{fig:recording-setup-3}).
In some cases (company~\textsf{B}), the same machine would be used for both
development and recording
(see \cref{fig:recording-setup-1})
thus limiting the available resources available in the developers' environment.

For the audio channel, there were also two solutions.
The easy one was to use the webcam microphone
(companies \textsf{A} and \textsf{B},
\cref{fig:recording-setup-1,fig:recording-setup-2}),
which was not able to pick up the developers' speech independently making it
difficult to understand them when both speak at the same time.
For higher audio quality, two wireless microphones were used in later
recordings (companies \textsf{C} to \textsf{F}, \cref{fig:recording-setup-3}).
Using an external USB soundcard, the two mono channels were hard-panned (100\%
left and 100\% right) and mixed to one stereo channel.\footnote{%
    Model of the microphones: Audio-Technica W-701/L;
    Soundcard: Tascam US-122L
}

\subsection*{Generation 2}

For the recordings in companies \textsf{J} and \textsf{K}, we used the web
conferencing tool Adobe Connect\footnote{%
    Product homepage: \url{https://www.adobe.com/products/adobeconnect.html}
}
which relies on a Flash-based browser plugin to share one's screen
content, webcam, and microphone.

To record the distributed pair programming sessions of company~\textsf{J},
two Adobe connect meetings were started such that both
developers each shared their respective screen, webcam, and headset microphone
(which they had to use anyway for their \hyperref[mode:dpp]{DPP} session).
On the receiving end, the researcher stacks the two screen outputs on top of
each other on a vertically rotated monitor and records the whole configuration
(see \cref{fig:recording-setup-4}).
Schenk organized the setup and the recording of
the \textsf{J}-sessions; see \cite[Sec.~3.2.2 \& 3.2.3]{Schenk18} for more details.

In companies~\textsf{K} and \textsf{M} a similar setup was used:
The developers put the headbands of USB headsets in the nape of their necks
such that the cushions do not cover their ears and the microphone can pick up
their speech.
Again, two Adobe Connect meetings were necessary, this time to transmit both
microphone signals.
In company \textsf{K}, the pairs' dual-screen setup could also be captured this
way (see \cref{fig:recording-setup-5}).

Compared to the first generation, the advantages of this infrastructure were
(a)~that the session could be watched live by the researcher while it was still
recording thus making the next step (quick analysis) a bit faster and
(b)~the ease of the setup allowed that the recording could be done
completely remotely as the developers had all necessary equipment on-site
(modern web-browser, webcam, and headsets).
Disadvantages were that
(c)~this setup only worked for Windows machines and
that support for Flash (necessary for running Adobe Connect) was already
declining at the time, and
(d)~fluctuation in network latency for the two audio channels meant that, for
co-located pair programming in companies \textsf{K} and \textsf{M},
there was a notable randomly changing offset between the two signals.
Whenever both microphones picked up the sound of one developer, e.g., because
they looked at each other, an annoying robotic echo could be the result, which
makes listening to the record rather unpleasant.

\subsection*{Generation 3}

In order to enable recordings of pair programming sessions independent of the
developers' operating systems, we looked for other solutions than Adobe connect.
The most versatile ones turned out to be Skype\footnote{%
    Product homepage: \url{https://www.skype.com}
} and TeamViewer,\footnote{%
    Product homepage: \url{https://www.teamviewer.com}
} the latter of which was able to transmit and record pixel-accurate
screen contents so our research group purchased a commercial license.
TeamViewer's recording feature allows to losslessly record the contents of a,
say, Full HD display even if the recording device has a much smaller resolution
available such as a window on a Laptop screen
(see \cref{fig:recording-setup-6}).
In order to process TeamViewer data further, however, the recording first
needed to be rendered as a pixel video, which at times took much longer than the
recording itself.

\begin{figure*}
    \centering
    \begin{tikzpicture}
    \node at (0,0)
    {\includegraphics[width=14.5cm]{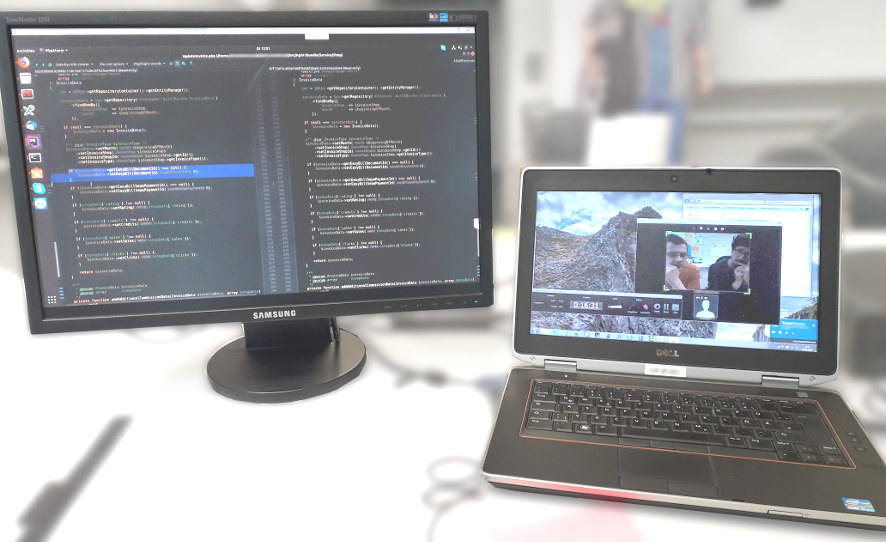}};

    \draw [green, ultra thick] (3.65,0.65) node (a) {}
    -- ++(1.4,-0.05) node (b) {}
    -- ++(-0.03,-0.97)
    -- ++(-1.4,0.07)
    -- cycle;

    \node at ($(a)!0.5!(b)$)
    [pin={[pin distance=1.6cm, pin edge={very thick, green}, align=center,
        fill=white!91!black]90:Recording webcam feed\\ with Camtasia}] {};

    \draw [green, ultra thick] (-7.05,3.65)
    -- ++(7.6,0.1)
    -- ++(0.05,-3.95) node (c) {}
    -- ++(-7.13,-0.4) node (d) {}
    -- cycle;

    \node at ($(c)!0.5!(d)$)
    [pin={[pin distance=1.8cm, pin edge={very thick, green}, align=center,
        fill=white!91!black]270:Recording screencast\\ with TeamViewer}] {};
    \end{tikzpicture}
    \caption{%
        Concurrent recording of screencast and webcam as shown in
        \cref{fig:recording-setup-6}.
        This is an annotated photograph taken during the recording of session
        \sRef{PA2} in a conference room at company \textsf{P}.
        The developers P1 and P2 sit two rooms further down the hall.
        We used an external monitor only to follow the session live;
        it would not have been strictly necessary because
        TeamViewer will record the screencast in full resolution regardless of
        its current display size.}
    \label{fig:screen-cam-parallel}
\end{figure*}

There were two options for recording the webcam.
For relatively small distances between the development and the recording
computers, the webcam could be connected directly to the recording machine
using a really long USB cable in order to not disturb the developers.
For setups more than one room apart, the webcam feed was transmitted in the
TeamViewer session, which posed additional problems:
Even though the webcam video is included in the pixel-video export, it has a
fixed position and small resolution
(ca.\ 160x120 pixels, next to the screen cast, in the upper right corner),
and is also often notably out of sync with the screencast.
To compensate this, we used Camtasia to screencapture the webcam feed separately
while the session is running at a resolution of about 460x340
(see schematic in \cref{fig:recording-setup-6}, and annotated photograph from
recording on-site in \cref{fig:screen-cam-parallel}).

To avoid the robotic echo of network-transmitted dual audio, we record the
audio of co-located sessions locally without any network in between.
In company~\textsf{O}, we used the same wireless microphones as before and fed
them into one dictaphone;
in company~\textsf{P} we used two independent dictaphones each wired to a simple
lapel microphone to record the developers individually.\footnote{%
    Dictaphone model: Olympus VN-8700PC;
    Microphone model: AV-JEFE TCM 141
}

\subsection*{Synchronizing the Data Channels}

Depending on the setup, a recording session would yield a number of files with
a number of data streams.
In the simplest form (as for companies \textsf{A} and \textsf{B}) the result
would be a single Camtasia file containing the screencast stream and the webcam
stream (which also includes the audio).
Such a file could be opened directly in Camtasia Studio and exported to a
self-contained video file such as an AVI or MP4 video file.

In companies \textsf{C} through \textsf{F}---where audio was recorded
externally---there were two Camtasia files:
One comprising the screen content and webcam video just as in the simplest
setup;
the other containing the externally recorded dual-mono-mixed-to-stereo audio
signal.
To ease synchronizing the separate audio and video channels, the developers in
these recordings were asked to clap in front of the camera.
Ultimately, there were two `timelines' that needed to be aligned.

With the second generation recording infrastructure using Adobe Connect web
conferences, no such alignment was strictly necessary as all individual streams
(two screens, one or two webcams, two audio channels) were already synchronized
the moment they were captured off the researcher's screen.
However, some audio post-processing was still necessary as Adobe Connect
center-panned both audio channels during the meeting such that they were mixed
together in the Camtasia recording.
Luckily, Adobe Connect allowed to record the sessions on the conference
server and provided a large ZIP file containing all individual streams,
including the audio track as an MP3 file.
Using the mixed audio from the Camtasia recording as a reference, we could align
the two separate audio recordings and hard-pan them to ultimately get a
self-contained video after aligning these three timelines.

In the third recording generation, which relies on TeamViewer, a total of
\emph{four} timelines need to be synchronized:
The screencast;
two independent dictaphone recordings which, even though they come from the
same model with identical settings, differ inexplicably in recording speed by
several seconds per hour;
and the webcam feed.
To synchronize these, we first stretched one of the audio recordings
(without changing the pitch) to match the other,
then looked for isolated code changes in the screencast and listen for
audible keystrokes,
then noticed that the either the screencast or audio needed to be stretched
again as they, too, drifted further and further apart towards the end of the
session,
and finally did the same again for the webcam video (looking for bilabial
plosives in the developers' speech, such as a [p] or [b])
which sometimes also needed some stretching by yet a different factor.

For a reproducible production from the raw material involving the various
synchronization steps, we use the open-source video scripting system
AviSynth.\footnote{%
    Homepage: \url{http://avisynth.nl/index.php/Main_Page}
}
An AviSynth script is executed by the AviSynth frame server which then serves
compatible programs such VirtualDub\footnote{%
    Homepage: \url{http://www.virtualdub.org}
} with video frames and audio samples.
These can either be played back in realtime (at least for simple scripts) or
be sent to an encoder such as x264\footnote{%
    Homepage: \url{https://www.videolan.org/developers/x264.html}}
to obtain a self-contained video file.
Apart from being reproducible, video production through AviSynth did not put
constraints on the video dimensions (that is, no constraints on top of the
H.264 standard)---unlike Camtasia Studio which up to and
including version 8 only supported a maximum resolution of 2048 by 2048 pixels.

Overall, the production of a single session video in the last setup would take
at least one day of work with many iterations of manual finetuning.

\end{document}